\documentclass[aip,jmp,reprint,amsmath,amssymb,superscriptaddress,longbibliography,eqsecnum,nofootinbib]{revtex4-2}
\usepackage{graphicx,bm,color,mathptmx}
\usepackage{}
\usepackage{hyperref}
\hypersetup{colorlinks=true,linkcolor=blue,citecolor=blue,urlcolor=blue,bookmarks=false}
\newcommand{\Tr}{\mathop{\mathrm{Tr}} \nolimits}

\newcommand{\sugg}[1]{#1}
\newcommand{\op}[1]{\hat{#1}}

\newcommand{\Var}{\mathop{\mathrm{Var}}\nolimits}
\newcommand{\Cov}{\mathop{\mathrm{Cov}} \nolimits}

\newcommand{\Fish}{\mathop{\mathsf{F}} \nolimits}

\begin{document}

\title{Extremal quantum states}

\author{Aaron~Z.~Goldberg}
\affiliation{Department of Physics, University of Toronto, Toronto, 
Ontario M5S 1A7, Canada}

\author{Andrei~B.~Klimov}
\affiliation{Departamento de Fisica, Universidad de Guadalajara, 44420~Guadalajara, Jalisco, Mexico}

\author{Markus~Grassl}
\affiliation{Max-Planck-Institut f\"ur die Physik des Lichts, 
  Staudtstra\ss e 2, 91058 Erlangen, Germany}
\affiliation{International Centre for Theory of Quantum Technologies, University of Gda\'{n}sk, 80~308 Gda\'{n}sk, Poland}

\author{Gerd~Leuchs}
\affiliation{Max-Planck-Institut f\"ur die Physik des Lichts, 
  Staudtstra\ss e 2, 91058 Erlangen, Germany}
\affiliation{Institute of Applied Physics, Russian Academy of Sciences, 603950~ Nizhny Novgorod, Russia}

\author{Luis~L.~S\'anchez-Soto}
\affiliation{Max-Planck-Institut f\"ur die Physik des Lichts, 
  Staudtstra\ss e 2, 91058 Erlangen, Germany}
\affiliation{Departamento de  \'Optica, Facultad de F\'{\i}sica, 
   Universidad Complutense, 28040~Madrid, Spain}

\begin{abstract}
The striking differences between quantum and classical systems predicate disruptive quantum technologies. We peruse quantumness from a variety of viewpoints, concentrating on phase-space formulations because they can be applied beyond particular symmetry groups. The symmetry-transcending properties of the Husimi $Q$ function make it our basic tool. In terms of the latter, we examine quantities such as the Wehrl entropy, inverse participation ratio, cumulative multipolar distribution, and metrological power, which are linked to intrinsic properties of any quantum state. We use these quantities to formulate extremal principles and determine in this way which states are the most and least ``quantum;'' the corresponding properties and potential usefulness of each \sugg{extremal} principle are explored in detail. \sugg{While the extrema largely coincide for continuous-variable systems, our analysis of spin systems shows that care must be taken when applying an extremal principle to new contexts.}
\end{abstract}
\maketitle

\tableofcontents

\section{Introduction}

Despite heated foundational debate concerning its interpretation~\cite{Pusey:2012aa}, it seems indisputable that the notion of a quantum state is a keystone of quantum theory~\cite{Peres:2002oz}.  Quantum states are ultimately responsible for the various counterintuitive features that cannot be fathomed in classical scenarios: these include quantum superpositions, coherence, no-cloning, entanglement, and quantum correlations, to cite but a few~\cite{Haroche:2006aa}. The hunt for useful quantum technologies in fields as diverse as computing, metrology, and information processing relies precisely on exploiting these remarkable resources.  

A natural question arises: how quantum is a state? What makes one state more quantum than another? Is there a trait to which one can point and say ``this embodies quantumness''? In this respect, we deem quantumness to be the most appropriate term to refer to the quantum nature of a state, rather than nonclassicality, which is quite often used and implicitly assumes a classical perspective.  

In quantum optics, quantumness is largely linked to the particle aspects of photons; whereas, photons' wave aspects can be well understood in classical terms. However, this is too heuristic to be of any quantitative use.  The quantumness of a system can be measured through a number of indicators, the most popular being squeezing~\cite{Lvovsky:2015aa,Loudon:1987aa,Andersen:2016aa}, sub-Poissonian statistics~\cite{Davidovich:1996aa}, and negativity of the Wigner function~\cite{Kenfack:2004aa}.  In terms of these indicators, numerous quantifiers have been discussed; relevant examples include Mandel’s $Q$ parameter~\cite{Mandel:1979aa,Mandel:1995qy}, distance-based measures~\cite{Hillery:1987aa,Dodonov:2000aa,Marian:2002aa,Dodonov:2003aa,Mari:2011aa,Nair:2017aa,Lemos:2018aa}, nonclassical depth~\cite{Lee:1991aa,Lutkenhaus:1995aa,Malbouisson:2003aa}, moment methods~\cite{Shchukin:2005aa,Gehrke:2012aa,Ryl:2015aa}, operator ordering sensitivity~\cite{Bievre:2019aa}, and resource-based indicators~\cite{Yadin:2018aa}. 
All these approaches have their own insights and merits in quantifying the quantumness of a state, but each highlights only one particular aspect of the problem. It seems impossible that there exists a unique and universal quantifier for quantumness suitable for all purposes. 

These quantifiers also have some drawbacks. First, most of them are bespoke, tailored to continuous variables or other special cases. For example, the established negativity of the Wigner function is merely a proxy for the non-Gaussianity of the state expressed in canonical position-momentum coordinates; this has no significance for other dynamical variables, such as the ones associated with spins and orbital angular momentum, for which the concept of Gaussianity is equivocal.  Second, many of them have no clear operational meaning and are hard to implement in the laboratory. This is the true, e.g., for the distance-based measures; they require a full tomography to be evaluated (which is challenging in many cases) and there are no physical arguments that favour any of the multiple metrics studied in the literature. Third, in some cases the resulting degrees of quantumness order states differently, from whence ensues, at the very least, puzzlement. \sugg{This is often related to the first drawback: in naively applying bespoke principles to other systems, one may be surprised that they disagree as to which states are the most quantum. We therefore advocate for the removal of a privileged example (such the harmonic oscillator) from a crowning of the most quantum states.}

In this paper we investigate indicators of quantumness that are of universal application and are not restricted to a particular symmetry.  The phase-space approach is germane to these purposes. Presenting quantum theory as a statistical theory on a classical phase space has attracted a lot of attention since the dawn of the former, providing clear insights about the classical-quantum boundary. In addition, this setting avoids the operator formalism, thereby freeing quantization from the burden of Hilbert space~\cite{Ali:2005aa}. Extensive literature on phase-space approaches to quantum theory can be found in books~\cite{Schroek:1996aa,Schleich:2001aa,QMPS:2005aa} and reviews~\cite{Hillery:1984aa,Lee:1995aa,Ozorio:1998aa,Polkovnikov:2010aa,Weinbub:2018aa}.

The phase-space approach maps the state of a system to a function on the phase space. However, this mapping, first suggested by Weyl~\cite{Weyl:1927aa} and later put on solid grounds by Stratonovitch~\cite{Stratonovich:1956aa}, is not unique: a whole family of $s$-parametrized quasiprobabilities can be assigned to each state. The most common choices of $s$ are $+1$, 0, and $-1$, which correspond to the $P$ (Glauber-Sudarshan)~\cite{Glauber:1963aa,Sudarshan:1963aa}, $W$~(Wigner)~\cite{Wigner:1932uq}, and $Q$~(Husimi-Kano)~\cite{Husimi:1940aa,Kano:1965aa} functions, respectively. For continuous variables, such as Cartesian position and momentum, the textbook example that triggered interest in this field, the parameter $s$ defines different orderings of the basic variables.

We stress that all of these quasiprobability functions contain complete information about the state; there is no privileged status for any of them. The choice of a particular instance depends exclusively on its convenience for the problem at hand. In this paper, we are unabashedly pro-Husimi, for it is positive and always exists regardless of \sugg{the choice of} phase space, revealing the quantumness of a state in a most conspicuous manner. \sugg{Moreover, its moments are straightforward to determine experimentally, allowing one to directly build information about a quantum state in a logical manner.}

More concretely, our aim is to use the Husimi $Q$-function as the basic ingredient to formulate sensible extremal principles, so we can determine in a consistent manner the least and the most quantum states.  Extremal principles are of paramount importance in physics, of which Fermat least-time, Hamilton least-action, and minimal entropy principles are well known examples. In some sense, extremal principles are guided by the Aristotelian notion of economy in nature. \sugg{In this respect, it is important to emphasize that even if the guiding quantities, such as action or entropy, sometimes have no simple operational definition, the resulting extremal states have complete physical significance.} Following this approach, we explore the states that are likely to play a major role in demonstrating true quantum advantages. 

A final word of caution: for composite systems,  quantum correlations~\cite{Adesso:2016aa} and entanglement measures~\cite{Horodecki:2009aa,Plenio:2019aa} provide powerful tools for inspecting quantumness. In addition, on physical grounds, the quantum properties are largely revealed through the projection associated with the measurement process~\cite{Korolkova:2019aa} and the concomitant modification of the state of the system. Therefore, to evaluate quantumness, one needs several sequential measurements on the same system. This calls for correlating two or more measurements: {then there are more subtle tests such as looking for variances of measurements, which concern the statistics of several independent single measurements}~\cite{Strekalov:2019aa}. These questions are certainly enthralling, but they lie outside the scope of our present work.

\section{Quantumness for continuous variables}

In the continuous-variable (CV) setting, quantum information is encoded in degrees of freedom with continuous spectra~\cite{Braunstein:2005aa,Andersen:2010aa}. The best-known example of this is the quantized harmonic oscillator, which can be described by position and momentum. The CV setting is an excellent tool to represent, among other examples, modes of the electromagnetic field, Josephson junctions, and Bose-Einstein condensates. In this section we present our quantumness indicators for this outstanding case.

\subsection{Canonical coherent states}

We start by briefly recalling the structures needed to set up a proper phase-space description of the Cartesian quantum mechanics of a harmonic oscillator. For simplicity, we choose one degree of freedom, so the associated phase space is the complex plane $\mathbb{C}$. The relevant observables are the Hermitian coordinate and momentum operators $\op{x}$ and $\op{p}$, with canonical commutation relation ($\hbar = 1$ throughout)
\begin{equation}
  \label{eq:HWcom}
  [\op{x}, \op{p}] = i \, \op{\openone} \, .
\end{equation}
These operators are the generators of the Heisenberg-Weyl algebra~\cite{Binz:2008oq}, which is the hallmark of noncommutativity in quantum theory. In general, the elements of this algebra can be written as linear combinations of the three generators; i.e., $i s \op{\openone} + i (p \op{x} - x \op{p})$. The elements of the group corresponding to the Lie algebra (\ref{eq:HWcom}) are obtained, as usual, by exponentiation: $\op{g}(s, x,p) = e^{i s}  \exp[i(p \op{x} -  x \op{p})]$. To interpret the action of this element, we recall that the action of the basic subgroups $\op{U} (x) = \exp (-i x \, \op{p})$ and $\op{V} (p) = \exp (i p \, \op{x})$ in the bases of eigenvectors of position and momentum is
\begin{equation}
  \label{eq:actbas}
  \op{U} (x^{\prime} ) | x \rangle = | x + x^{\prime} \rangle  \, ,
  \qquad 
  \op{V} (p^{\prime} ) | p \rangle = | p + p^{\prime} \rangle \, ,
\end{equation}
so they represent displacements along the corresponding coordinate axes. Therefore, the operator $\op{D}(x,p) = e^{ixp} \, \op{U}(x) \op{V}(p)$ corresponds to a displacement operator in phase space. 

Instead of the variables $\op{x}$ and $\op{p}$ another pair of operators is often more suitable: they are defined by
\begin{equation}
\op{a} = \frac{1}{\sqrt{2}} (\op{x} + i \op{p}) \, , 
\qquad \qquad
\op{a}^{\dagger} = \frac{1}{\sqrt{2}} (\op{x} - i \op{p}) \, , 
\end{equation}
so that their commutation relation is
\begin{equation}
[ \op{a}, \op{a}^{\dagger}] = \op{\openone} \, .
\end{equation}
The displacement operator can then be written as
\begin{equation}
  \op{D}( \alpha )  = \exp(\alpha \op{a}^\dagger - \alpha^{\ast} \op{a} ) \, ,
  \label{eq:displal}
\end{equation}
where the complex number $\alpha$ is $\alpha = (x + i p)/\sqrt{2}$.  
Note that $\op{D}^\dagger (\alpha ) = \op{D} (- \alpha)$, while
$\op{D}(0) = \op{\openone}$. Displacement operators obey the simple composition law
\begin{equation}
  \label{eq:com} 
  \op{D} (\alpha ) \, \op{D} (\beta ) = 
  \exp[\tfrac{1}{2} (\alpha \beta^{\ast} - \alpha^{\ast} \beta)] \, 
  \op{D} (\alpha + \beta ) \, .
\end{equation}

The displacements constitute a basic ingredient for the definition of coherent states. Let us choose a fixed, normalized reference state $ | \Psi_{0}\rangle $. For any $\alpha \in \mathbb{C}$, we can define the coherent state $| \alpha \rangle$ as~\cite{Perelomov:1986ly,Gazeau:2009aa}
\begin{equation}
  | \alpha \rangle = \op{D} ( \alpha ) \, | \Psi_{0} \rangle \, ,  
  \label{eq:defCS}
\end{equation}
so that the resulting set of states are parametrized by phase-space points and inherit properties from those of $\op{D} (\alpha)$.  The standard choice for the fiducial vector $| \Psi_{0} \rangle$ is the vacuum $|0 \rangle $. This has quite a number of relevant properties, which make any coherent state a minimum-uncertainty state; namely,
\begin{equation}
   \Var_{\alpha} (\op{x}) \, \Var_{\alpha}(\op{p} ) = \frac{1}{4} \, ,  
  \label{eq:MUS}
\end{equation}
where $\Var_{\varrho} (\op{A}) = \langle \op{A}^{2} \rangle_{\op{\varrho}} - \langle \op{A} \rangle_{\op{\varrho}}^{2}$ is the variance of the operator $\op{A}$.

Coherent states are not mutually orthogonal; their overlap is 
\begin{equation}
\lvert \langle \beta | \alpha \rangle \rvert^2 = \exp ( - |\alpha  - \beta
|^{2} ) \, , 
\end{equation}
but they can be considered as approximately orthogonal for sufficiently different values of $\alpha$ and $\beta$. Still, coherent states form an overcomplete basis in the space of states, so we have the resolution of unity
\begin{equation}
\frac{1}{\pi} \int_{\mathbb{C}}  d\alpha \; |\alpha \rangle \langle \alpha |  = \op{\openone} \
\end{equation}   
in terms of nonorthogonal projectors. This completeness relation allows us to expand an arbitrary pure state $| \psi \rangle$ as
\begin{equation}
|\psi \rangle = \frac{1}{\pi} \int_{\mathbb{C}} d\alpha \; 
\psi (\alpha^{\ast}) \, | \alpha \rangle  \, ,
\end{equation}
where $\psi (\alpha^{\ast}) = \langle \alpha | \psi \rangle$ is the wave function in the coherent-state representation, which completely determines the state $| \psi \rangle$.  If we take into account that the expansion on the Fock basis $\{ |n \rangle \}$ of a coherent state reads
\begin{equation}
| \alpha \rangle = e^{| \alpha |^{2}/2} 
\sum_{n=0}^{\infty} \frac{\alpha^{n}}{\sqrt{n!}} \, | n \rangle ,
\end{equation} 
we can write 
\begin{equation}
\psi ( \alpha ) = e^{- | \alpha |^{2}/2} f_{\psi} ( \alpha) \,, 
\qquad \qquad
f_{\psi} (\alpha) = \sum_{n=0}^{\infty} \frac{\psi_{n}}{\sqrt{n!}} 
\alpha^{n} \,, 
\end{equation}
where $f_{\psi} (\alpha) $ is an entire analytic function called the Bargmann-Segal representation of the state~\cite{Bargmann:1961aa,Segal:1978aa}.

Finally, we mention that coherent states are quite often singled out as quasi-classical, in the sense that the mean values of the position and momentum operators oscillate in agreement with classical laws. However, this is misleading because, according to the Ehrenfest theorem~\cite{Ehrenfest:1927aa}, this holds true for \emph{any} state under the dynamics of the harmonic oscillator.

\subsection{Husimi function and stellar representation}

The attempts to find a description of quantum states that are similar to classical states have given rise to several quasiprobability distributions, as discussed in the Introduction. 

The Wigner function permits a direct comparison between classical and quantum dynamics. For coherent and squeezed states, the Wigner function takes the form of a Gaussian. Generally, however, it is not a positive function and thus not a probability distribution, even though it is normalized. To overcome this shortcoming the Husimi $Q$-function is defined as
\begin{equation}
Q_{\varrho} ( \alpha ) = 
\Tr ( \op{\varrho} \, |\alpha \rangle \langle \alpha |) =
\langle \alpha | \op{\varrho} | \alpha \rangle \, .   
\end{equation}
It is everywhere nonnegative and properly normalized 
\begin{equation}
\frac{1}{\pi} \int_{\mathbb{C}} d\alpha \, 
Q_{\varrho} (\alpha )   = 1 \, , 
\end{equation}
and can therefore be interpreted as a genuine probability distribution. Nevertheless, unlike classical probability distribitutions, which obey no further constraints, it is also bounded from above by $1/\pi$. In addition, it is a distribution of a somewhat peculiar kind, since it is not a probability density for mutually exclusive events. Instead $Q_{\varrho} (\alpha)$ is the probability that the system, if measured, would be found in a coherent state whose probability density has its mean at $\alpha$. Such events are not mutually exclusive because the coherent states overlap with each other~\cite{Stenholm:1992aa}. 

Compared to the Wigner function, the $Q$-function has the disadvantage that one does not recover the correct marginals. Moreover, its definition depends on the definition of the coherent states, and hence on some special reference state in the Hilbert space. Yet, this oddity can be turned to an advantage. The $Q$-function is the pertinent quasiprobability distribution to employ when the measurement device introduces noise that can be modelled by the reference state used to define the $Q$-function~\cite{Leonhardt:1997aa}.

For a pure state $|\psi \rangle$, the $Q$-function reduces to
\begin{equation}
Q_{\psi} ( \alpha ) = | \langle \alpha | \psi \rangle |^{2} =
e^{- |\alpha |^{2}} \, |f_{\psi} ({\alpha^{\ast}})|^{2} \, .
\end{equation}
The zeros of the Husimi $Q$ function are then the complex conjugates of the zeros of the Bargmann function $f_{\psi}$: they constitute the so-called stellar representation of the state~\cite{Bengtsson:2006aa} (also called the Majorana constellation~\cite{Majorana:1932ul}). Two states with the  same constellation are the same, up to a global phase. {The function $f_\psi (\alpha)$ can thus be aptly called} the stellar function~\cite{Wehrl:1978aa}. Since this function is an analytical function, its zeros form a discrete set~\cite{Markushevich:2001aa}. 

Any single-mode Gaussian pure state may be obtained from the vacuum by way of a displacement $\op{D} (\beta) = \exp (\beta \op{a}^{\dagger} - \beta^{\ast} \op{a})$ and a squeezing operation $\op{S} (\xi) = \exp [ \tfrac{1}{2} (\xi \op{a}^{2} - \xi^{\ast} \op{a}^{\dagger 2})]$. The resulting state $|\beta, \xi \rangle$ is the most general {Gaussian pure} state and its corresponding stellar function is~\cite{Chabaud:2020aa}
\begin{equation}
f_{\beta, \xi} ( \alpha ) = ( 1 - | A |^{2} )^{1/4} \,
\exp \left ( - \tfrac{1}{2} A \alpha^{2} + B \alpha +  C \right )
\label{eq:Gaussian stellar},
\end{equation}
where 
\begin{equation}
A = e^{- i \theta } \tanh r \,, 
\qquad
B = \beta \sqrt{1- | A |^{2}} \, ,
\quad 
C = \tfrac{1}{2} ( A^{\ast} \beta^{2} - |\beta |^{2} ) \,  ,
\end{equation}
and $\xi = r \,e^{i \theta}$ {(with $r \ge 0$)} is the squeezing parameter. Setting $\xi =0$ yields the stellar function for the coherent state $| \beta \rangle$; whereas, setting $\beta = 0 $ yields the stellar function for the squeezed vacuum $| \xi \rangle $. It is clear that the stellar function of a Gaussian state does not have zeros and the associated constellation is void.  

At the opposite extreme, we have the number states,  with stellar functions
\begin{equation}
f_{n} (\alpha ) = \frac{\alpha^{n}}{\sqrt{n!}},
\end{equation}
so the constellations reduce to the origin, with  multiplicity $n$.

We can now use antistereographic projection to represent the roots as points on the unit sphere \sugg{whose south pole touches the origin of the complex plane}. It is straightforward to prove geometrically that the line that connects the north pole with a point $\alpha = \tan (\theta/2) \exp (i \phi)$ in the complex plane will intersect the unit sphere at a point with spherical coordinates $(\theta, \phi)$; this point is the associated Majorana star. In this picture, Gaussian states collapse to the north pole, while the south pole corresponds to number states. In between, we have states represented by a set of points on the sphere. As an intermediate example, we consider the cat states
\begin{equation}
| \mathrm{cat}_{\pm} \rangle = {\frac{1}{\sqrt{\mathcal{N}_{\pm}}}}
(|\beta \rangle \pm | - \beta \rangle) 
\label{eq:catpm}
\end{equation}
with ${\mathcal{N}_{\pm}} = 2 [ 1 \pm \exp ( -2 |\beta|^{2} )]$, {which have also been called even (+) and odd ($-$) coherent states~\cite{Dodonov:1974aa}.} For the {odd case} $| \mathrm{cat}_{-}\rangle$ and $\beta = i y$, we have
\begin{equation}
f_{\mathrm{cat}_{-}} (\alpha ) \propto \sin (\alpha y)
\end{equation}
and the constellation is now discrete but with infinite points. One can also find core states, as termed in Ref.~\onlinecite{Menzies:2009aa}, which are single-mode normalized states that have a polynomial stellar function (and so a discrete constellation). This evidence seems to suggest that quantumness can be linked to the geometrical properties of a state's constellation.

\subsection{Moments of the Husimi function}
\label{sec:moments}

The moments of $\op{x}$ and $\op{p}$ are often the most experimentally-accessible pieces of information.  For Gaussian states, the first two moments suffice, but, in general, higher-order moments contribute valuable information. Since the dynamical variables are noncommuting, higher-order moments make unambiguous sense only after we commit to a particular operator ordering scheme or, equivalently, to a particular quasiprobability. The case of Weyl (or symmetric) ordering, corresponding to the Wigner function, has been carefully examined~\cite{Ivan:2012aa}; here we solve the problem from the Husimi perspective. 

To elucidate the question we first briefly recall the so-called $s$-parametrized Stratonovitch-Weyl map~\cite{Gosson:2016aa}
\begin{equation}
  \op{A}  \mapsto  W_{A}^{(s)} (\alpha ) =
  \Tr [  \op{A} \, \op{w}^{(s)}(\alpha ) ]  \, ,
  \label{map}
\end{equation}
which maps each obervable $\op{A}$ onto a function on the complex plane $\mathbb{C}$. The corresponding kernels $\op{w}^{(s)}$ were worked out by Cahill and Glauber~\cite{Cahill:1969aa}:
\begin{equation}
  \label{eq:wCG}
  \op{w}^{(s)} (\alpha) = \frac{1}{\pi}
  \int_{\mathbb{C}} d \beta \, 
  e^{s |\beta|^{2}/2} \, \exp(\alpha \beta^\ast - \alpha^\ast \beta)
  \, \op{D} (\beta) \,  .
\end{equation}   
The function $W_{A}^{(s)} (\alpha)$ is called the symbol of the operator.  The value $s=0$ corresponds to the Weyl symbol (symmetric ordering); whereas, $s =1$ and $s =-1$ lead to the contravariant $P$-symbol (normal ordering) and covariant $Q$-symbol (antinormal ordering), respectively. In particular, the symbols of the density operator are the traditional quasiprobability distributions: they are all covariant under displacements and provide a basic overlap relation
\begin{equation} 
\Tr (\op{\varrho} \, A) = \frac{1}{\pi} \int_{\mathbb{C}} d\alpha \, 
W_{\varrho}^{(s)} (\alpha) W_{A}^{(-s)} (\alpha ) \, .
\end{equation}
From this viewpoint, the Wigner symbol is self-dual, while the $P$ and $Q$ symbols are dual to each other. To stress this point, we will use the notation $\op{w}^{(+1)} (\alpha) = \op{\mathfrak{p}}(\alpha )$  and 
$\op{w}^{(-1)} =  \op{\mathfrak{q}} (\alpha )$, so that
\begin{equation}
Q( \alpha ) = \Tr [ \op{\varrho} \, \op{\mathfrak{p}} ( \alpha)] \, ,
\qquad \qquad
P( \alpha ) = \Tr [ \op{\varrho} \, \op{\mathfrak{q}} ( \alpha)] \, .
\end{equation}

Using the definition \eqref{eq:wCG}, the $\op{\mathfrak{q}}$-kernel can be explicitly written as%
\begin{equation}
\op{\mathfrak{q}} (\alpha ) = 
\sum_{n,m} \mathfrak{q}_{nm} (\alpha ) \; \op{a}^{\dagger n} \op{a}^{m},
\label{q2}
\end{equation}%
{with}
\begin{equation}
\mathfrak{q}_{nm}(\alpha ) = \left 
\{ 
\begin{array}{c}
\frac{(-1)^{m}}{n!} \alpha^{n-m}e^{-|\alpha |^{2}}
L_{m}^{(n-m)} ( |\alpha |^{2} ) ,\qquad n \geq m \\ 
\\
\\
\frac{(-1)^{n}}{m!}\alpha ^{\ast m-n}e^{-|\alpha |^{2}}
L_{n}^{(m-n)} ( |\alpha |^{2} ) ,\quad m\geq n%
\end{array}%
\right. ,
\end{equation}
\sugg{where $L_{n}^{(k)} (x)$ are the associated Laguerre polynomials.}
Equation~(\ref{q2}) can be rewritten in the following compact form:
\begin{equation}
\op{\mathfrak{q}}(\alpha ) = \sum_{K} 
\sum_{q=-K}^{K} \mathfrak{q}_{K+q \; K-q}(\alpha ) \;
\op{T}_{Kq},  
\end{equation}
where $K =0, 1/2, 1, \ldots$ and we have introduced the operators
\begin{equation}
\op{T}_{Kq}  = \op{a}^{\dagger K+q} \op{a}^{K-q} \, ,
\end{equation}
which constitute a basis in the algebra of monomials. These homogeneous polynomials form a set and transform linearly among themselves under symplectic transformations in Sp(2, $\mathbb{R}$); i.e., transformations that preserve the Poisson bracket for classical descriptions or the commutator \eqref{eq:HWcom} for quantum ones \cite{Guillemin:1984aa,Gosson:2000aa}. In this way, we can expand an arbitrary density operator as
\begin{equation}
\op{\varrho} = \sum_{K} \sum_{q=-K}^{K} \varrho_{Kq} \op{T}_{Kq},
\end{equation} 
where the coefficients can be read off from 
\begin{equation}
\label{eq:pqsi}
\varrho_{Kq} = \frac{1}{\pi} \int_{\mathbb{C}} d\alpha \, P(\alpha ) \; 
\mathfrak{q}_{K+q \, K-q}(\alpha ) \, .
\end{equation}%
These coefficients can be called the state multipoles, in close analogy with the established techniques that we will use in Sec.\ref{sec:multipoles}.

The nonnegativity of the density operator of a state is faithfully encoded in the Husimi $Q$ function, which then constrains the moments of the Husimi distribution. These constraints appear in a canonically covariant form when expressed in terms of the multipoles. Actually, the cumulative multipole distribution 
\begin{equation}
\mathcal{A}_{M} = \sum_{\sugg{K=1/2}}^{M} \sum_{q=-K}^{K} |\varrho _{Kq}|^{2} \, ,
\end{equation}
can be interpreted as a generalized uncertainty principle~\cite{Ivan:2012aa}. {This will play a crucial role in Sec.~\ref{sec:spins} as a good indicator of quantumness. It is reasonable that coherent states maximize the quantity $\mathcal{A}_{M}$, at least for large $M$ and, again, are the least quantum states.}

{The states which minimize $\mathcal{A}_{M}$ can be seen as the most quantum ones. For a spin, they will be called the Kings of Quantumness. We thus look for the constraints on the density operator $\op{\varrho}$ that allow the cumulative multipoles to vanish to a given order. This imposes that all $\varrho_{Kq}=0$ for $K\leq M$ for some $M$.}

{We first examine the case $q=0$. According to \eqref{eq:pqsi} we have }
\begin{equation}
{\varrho_{K0} = \int d\alpha \; P(\alpha) \, \mathfrak{q}_{KK}(\alpha) \propto 
\int d\alpha \; P(\alpha) e^{-|\alpha|^2} \, L_{K}(|\alpha|^2)\, ,}
\end{equation}
{where $L_{K} (x)$ are the Laguerre polynomials. \sugg{The function} $P(\alpha)$ can be formally expressed as~\cite{Perina:1991aa}}
\begin{equation}
{P(\alpha) = \sum_{n,m} (-1)^{n+m} \varrho_{nm} \frac{\sqrt{m!n!}}{2\pi r(n+m)!} 
e^{r^2} e^{i(m-n) \theta} \left(\frac{\partial}{\partial r}\right)^{m+n}\delta(r) \, ,}
\end{equation}
{where $\rho_{n,m}$ are the matrix elements of the density operator in the Fock basis and we have expressed the complex amplitude in polar coordinates $\alpha = r \exp (i \theta)$. Since $\mathfrak{q}_{KK}(\alpha)=\mathfrak{q}_{KK}(r)$ the integral over $\theta$ is readily carried out. The radial part can be integrated by parts, with the result}
\begin{equation}
{\varrho_{K0} \propto \frac{1}{2}\sum_{n}\rho_{n,n}\frac{n!}{(2n)!} 
\left(\frac{\partial}{\partial r}\right)^{2n}L_K (r^2)\bigg|_{r=0} = \frac{1}{2}
\sum_{n=0}^K  (-1)^n \rho_{n,n} \binom{K}{n} \, .}
\end{equation}  
{For $\varrho_{K0}=0$ to hold for all integers $K\leq M$, we must have all $\rho_{n,n}$ be equal for $n\leq M$. If these components are nonzero then the state must be mixed.
All states whose diagonal components $\rho_{n,n}$ are only nonzero for $ n>K$ have $\varrho_{K0}=0$. The lowest-energy state with this property is clearly the Fock state $|K+1 \rangle$. This state is pure.}

{For $q\neq 0$ a similar calculation shows that}
\begin{equation}
\varrho_{Kq} \propto \sum_{n=0}^{K-q} (-1)^n \rho_{n+2q,n}
\frac{\sqrt{n!(n+2q)!}}{n!} \binom{K+q}{n+2q} \, ,  \qquad q >0 \, ,
\end{equation}
{and an analogous expression for $q < 0$. From here, one can check  that all of the off-diagonal components smaller than $M$ must vanish for a state to have $\mathcal{A}_{M} = 0$. Borrowing the terminology of spins, we refer to these states as $M$-th order upolarized.} 

{We thus conclude that to be unpolarized to order $M$, a state can either have no components with energy less than or equal to $M$ or can be proportional to the identity on that subspace. The lowest-energy pure state that is unpolarized to order $M$ is thus the lowest-energy pure state that only has components with more energy than $M$: the Fock state $| {M+1} \rangle$.}

\subsection{Wehrl entropy}

The standard von Neumann definition of the quantum-mechanical entropy 
\begin{equation}
{S_{Q} (\op{\varrho})} = - \Tr (\op{\varrho} \ln \op{\varrho} )
\end{equation} 
vanishes for all pure states. Thus, it cannot distinguish between various pure states and it is rather a measure of the purity of quantum states. A different definition of classical-like entropy associated with a quantum state {was} proposed by Wehrl~\cite{Wehrl:1978aa}
\begin{equation} 
S_{\mathrm{W}} (\op{\varrho}) = - \frac{1}{\pi} 
\int_{\mathbb{C}} d\alpha \;
Q_{\varrho} (\alpha) \,  \ln Q_{\varrho} (\alpha) \, .
\end{equation}
The Wehrl entropy can loosely be interpreted as an information measure for a joint noisy measurement of $x$ and $p$. The $Q$-function can never be so concentrated that $S_{\mathrm{W}}$ becomes negative. On the contrary, classical distributions can be arbitrarily concentrated in phase space and classical entropies can take on negative values~\cite{Orowski:1999aa}. They may even tend to minus infinity if their distributions tend to delta functions.  Moreover, Wehrl proved the even stronger relationship 
\begin{equation}
S_{\mathrm{W}} (\op{\varrho} ) \ge S_{Q} (\op{\varrho}) \, ,
\end{equation}
which establishes a connection between the Wehrl and the von Neumann entropies of a given state. 

However, the most important property of this classical-like entropy is, perhaps, the following inequality:
\begin{equation}
S_{\mathrm{W}} (\op{\varrho} ) \geq 1 \, .
\end{equation}
Lieb~\cite{Lieb:1978aa}, using sophisticated techniques from functional analysis, was able to demonstrate  that the equality holds if, and only if, the considered state is a coherent state. In other words, coherent states minimize the Wehrl entropy.  

We observe that the Wehrl entropy is a good measure of the strength of the coherent component. That is, it measures how “close” a given state is to a coherent state. For any coherent state the Wehrl entropy is identically equal to one. The Wehrl entropy does not change when a state undergoes a displacement operation: {for general Gaussian states, we have} 
\begin{equation}
S_{\mathrm{W}} (| \beta, \xi \rangle ) = 1 + \ln \cosh r \, .
\end{equation}
Thus, the Wehrl entropy of squeezed states with nonzero coherent component is smaller than the squeezed vacuum with the same average photon number. Observe that the Wehrl entropy is independent from the coherent component of the state.

The natural question is what are the states the maximize the Wehrl entropy. If we fix the average photon number $\bar{n} = \langle \op{N} \rangle$ a simple calculus of variations gives~\cite{Orowski:1999aa}
 \begin{equation}
Q_{\mathrm{W}}^{\sugg{\mathrm{max}}} ( \alpha ) = \frac{1}{\bar{n} + 1} \exp\left(\sugg{-} \frac{| \alpha |^{2}}{\bar{n} + 1} \right ) \, ,
\end{equation}
which corresponds to single-mode thermal radiation. However, this is a mixed state. A more fair way of proceeding would be to restrict ourselves to  pure states, as this is the case for coherent states. Our physical intuition suggests that we first assess number states; a direct calculation gives
\begin{equation}
S_{\mathrm{W}} ( |n\rangle ) = 1 + n + \ln n! - n \, \psi (n+1) \,, 
\end{equation}
where $\psi (n)$ is the digamma function~\cite{Abram:1996ul}. 

Interpolating between coherent states and number states are the photon-added coherent states~\cite{Agarwal:1991aa}
\begin{equation}
| \beta, m \rangle = \frac{1}{\sqrt{\mathcal{N}}} 
\; \op{a}^{\dagger\, m} | \beta \rangle \, , 
\qquad 
\mathcal{N} = m! \; L_{m} (- |\beta |^2 )  \, .
\end{equation} 
{These states offer the opportunity to closely follow the smooth transition between the particlelike and the wavelike behaviour of light~\cite{Parigi:2007aa}. Their mean number of photons is given by }
\begin{equation}
\bar{n}=
\frac{(m+1) L_{m+1} (- | \beta |^2 )}{L_{m} ( - |\beta |^2 )} - 1 \,,
\end{equation}
{which increases with $ |\beta |$ and $m$.} 

{The Wehrl entropy for these states can be analytically computed. The final result looks daunting; viz.,}
\begin{eqnarray}
S_{\mathrm{W}} (|\beta, m \rangle ) & = & 
\frac{e^{- |\beta |^2}}{{m! L_{m} (- |\beta|^)}} m! 
\Biggr \{ m \, _1F_1 (m + 1; 1; |\beta |^2 ) [1  - \psi(m+1) ] 
- m \, |\beta| ^2 \, _1F_1 (m+1; 2; |\beta|^2 ) \nonumber \\
& + & \ln [ m! \, L_{m} (- |\beta|^{2}) +1  - m 
\frac{\partial L_{-m-1}( |\beta | ^2 )}{\partial m} \Biggr \} \, ,
\end{eqnarray}
{where $ _1F_1 (a; b; z )$ is the Gauss confluent hypergeometric function.  However, we recover the proper limits: for small $|\beta |$, the Wehrl entropy becomes}
\begin{equation}
S_{\mathrm{W}} (|\beta, m \rangle ) \simeq 1 + m + \ln (m!) - m \psi (m +1) 
- |\beta | ^2 m \frac{\partial^2 L_{-m-1}(- x )}
{\partial m \; \partial x} \Biggr |_{x=0} ,
\end{equation} 
which decreases with $ |\beta |$; whereas, for large $|\beta |$, {$S_{\mathrm{W}} (|\beta, m \rangle )$ decreases} to 1.

\begin{figure}
    \centering
    \includegraphics{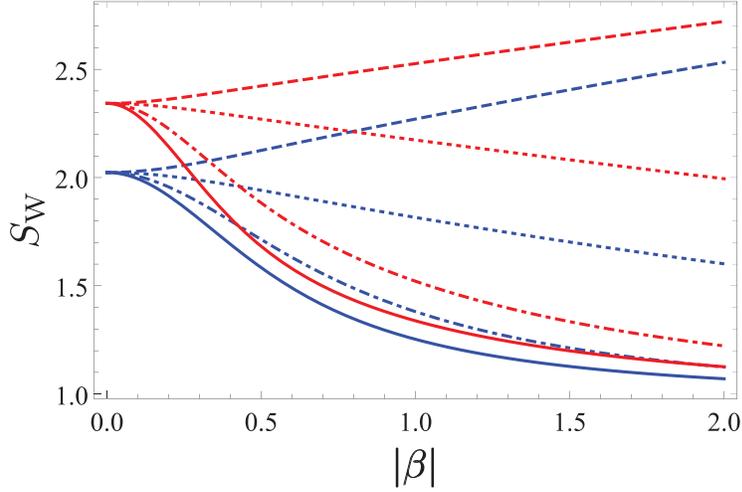}
    \caption{Wehrl enropies of photon-added coherent states $|\beta, m\rangle$ versus Fock states with the same average energy. The Wehrl entropies are the solid lines (blue is $m=3$ and red is $m=6$) and decrease monotonically with $ | \beta |$ between those of Fock states and those of coherent states. Fock states with the same average energy as these states would have entropies given by the dashed lines, which increase monotonically with $|\beta |$. Displacing these states lowers their energy; Fock states with the same average energy as the antidisplaced states $\op{D}^\dagger(\beta)|\beta, m\rangle$ (dotted lines) and as the optimally-displaced states $\op{D}^\dagger(\gamma)|\beta, m\rangle$ (dot-dashed lines) still have greater entropies than the photon-added coherent states.}
    \label{fig:photon added Wehrl}
\end{figure}

Even though displacing a state has no effect on its Wehrl entropy, it does have an effect on its average photon number \sugg{$\bar{n}$}. This means that there exist states with the same Wehrl entropy as $| \beta, m \rangle$, but with a smaller average number of photons. Displacing a Fock state always increases the average photon number; similarly, it is easy to find the $\gamma$ such that $\op{D}^\dagger(\gamma ) | \beta, m \rangle $ has minimal $\bar{n}$. Nonetheless, Fock states with this minimum average photon number still have greater Wehrl entropy than $| \beta, m \rangle$ (\sugg{see} Fig.\ref{fig:photon added Wehrl}). Similar considerations help motivate our conjecture that Fock states maximize the Wehrl entropy for a given average photon number.

\subsection{Inverse participation ratio}

The von Neumann entropy $S_{Q}$, containing the factor {$\ln \op{\varrho}$}, involves all the powers {of the density operator}, which complicates its evaluation. For this reason, it is useful to linearize the logarithm as $\ln \op{\varrho} = \ln [1 - (1- \op{\varrho})] \simeq {- (1 - \op{\varrho})}$ and then $S_{Q}$ reduces to $S_{L} = 1 - \Tr (\op{\varrho}^{2})$. The resulting linear entropy is just as good as the von Neumann entropy for quantifying the degree of purity of a state. But $S_{L}$ is easier to calculate than $S_{Q}$, given that it does not require diagonalization of the density matrix, hence its appeal for quantum information.

The same linearization can be applied to the Wehrl entropy. The resulting measure
\begin{equation}
\label{eq:SWlinCV}
S_{\mathrm{W, lin}} (\op{\varrho}) = 1 - {\frac{1}{\pi}}
\int_{\mathbb{C}} d\alpha \; {Q_{\varrho}}^{2} (\alpha )
\end{equation} 
might be called the linear Wehrl entropy.  Interestingly, the integral of the square of the Husimi distribution appearing in the definition of $S_{\mathrm{W, lin}}$,
\begin{equation}
M_{2} (\op{\varrho}) = {\frac{1}{\pi}} 
\int_{\mathbb{C}} d\alpha \; {Q_{\varrho}}^{2} (\alpha )\, ,
\end{equation}
does not reduce to the purity, as would happen if one would calculate the square of the Wigner function.~\footnote{Notice that, the Husimi $Q$ function being a probability density, its moments must be understood as in Sec.~\ref{sec:moments}. However, one often finds in the literature the term moment~\cite{Sugita:2002aa,Sugita:2003aa} referring to quantities like $M_{2}$.}  

For a pure state with amplitudes $\psi_{n}$ in the Fock basis,   $M_{2} (\op{\varrho} ) $ can be expressed as
\begin{equation}
M_{2} (\op{\varrho}) = \sum_{n, n^{\prime}} \sum_{m, m^{\prime}} 
\frac{\psi_{n} \psi_{n^{\prime}} \psi_{m}^{\ast} \psi_{m^{\prime}}^{\ast}}
{n! \, n^{\prime}! \; m! \, m^{\prime}!}
\int_{\mathbb{C}} d\alpha \; e^{- 2 | \alpha |^{2}} \alpha^{\ast n+n^{\prime}}
\alpha^{m + m^{\prime}} \, .
 \end{equation} 
The integral {can be directly computed to yield the compact result~\cite{Sugita:2002aa}}
\begin{equation}
M_{2} (\op{\varrho}) = \frac{1}{2} \sum_{K=0}^{\infty} | B_{K} |^{2} 
\end{equation}
with 
\begin{equation} 
B_{K} = \sum_{L=0}^{K} \sqrt{\frac{K!}{2^{K} \, L! (K- L)!}} 
\psi_{L} \psi_{K-L} \, .
\end{equation}

Explicitly related to $S_{\mathrm{W, lin}}$ is the quantity 
\begin{equation}
R(\op{\varrho}) = \frac{1}{M_{2} (\op{\varrho})} \, ,
\end{equation}
known as the inverse participation ratio, which has been widely used as a measure of the localization of a state in phase space~\cite{Gnutzmann:2001aa,Wobst:2003aa,Aulbach:2004aa}. 

For a coherent state $| \beta \rangle$, we have 
\begin{equation}
R (| \beta \rangle ) = 2 \, ,
\end{equation}
and this corresponds to the most localized state. Squeezing delocalizes the state
\begin{equation}
R ( |\beta,\xi \rangle) = {2 \cosh r
\simeq 2  \sqrt{\bar{n}}} \, ,
\qquad \qquad \bar{n}=\sinh^2 r \gg 1 \, .
\label{eq:M2squeezed}
\end{equation}
{From this perspective, quantumness is loosely associated with delocalization in phase space.

Again, we assess the question for number states: their only nonzero element is $| B_{K} |^2 = \delta_{K,2n} 2^{-K} \binom{K}{K/2}$ and, consequently,}
\begin{equation}
R( |{n} \rangle ) =  {2^{2n+1} \, \binom{2n}{n}^{-1} 
\simeq 2 {\sqrt{n \pi}}} 
, \qquad n \gg 1.
\end{equation}
This is the same scaling with $\bar{n}$ as for squeezed states \eqref{eq:M2squeezed}; {but number states are less localized by a factor of at least $\sqrt{\pi}$.}

{We can also look at intermediate cases, such as the states}
\begin{equation}
|{\delta} \rangle = \sqrt{p} |{n} \rangle +  \sqrt{1-p} |{n+\delta} \rangle \, \end{equation}
with average photon number $\bar{n}=n+(p-1)\delta$. {Their moments  decrease at the same order of $\bar{n}$ as squeezed and number states, but with a slower prefactor, regardless of the sign of $\delta$, which only contributes on the order of $\bar{n}^{-3/2}$}:
\begin{equation}
{R ( |\delta \rangle) \simeq \frac{2\sqrt{\bar{n}\pi}}{1+2p(1-p)}  \, ,
\qquad n \gg 1, \delta \, .}
\end{equation} 
The {participation ratio is smaller for $|\delta \rangle $ than for $|{n} \rangle$ for the simple reason that the former has more nonzero coefficients $\psi_n$. Since $1+2p(1-p) <\sqrt{\pi}$, the states $|\delta \rangle$ are less localized than squeezed states for all $p$.}

\subsection{Husimi extrema}
\label{sec:maxQ}

The inverse participation ratio measures the localization of the Husimi function through its second-order moment. Another {sensible} way of quantifying localization is the maximum value attained by the $Q$ function, which is a proxy for determining the infinite-order moment. Therefore, if we restrict for simplicity to pure states, we consider the quantity~\cite{Marian:2020aa}:
\begin{equation}
{M_{\infty}} (|{\psi}\rangle ) = \max_{\alpha} Q_\psi (\alpha).
\label{eq:geometric quantum Marian}
\end{equation} 
This quantity is a proper distance measure for pure states, and is invariant under arbitrary displacements. {It can be related to the Hilbert-Schmidt distance between a state $|\psi \rangle$ and the convex set of coherent states, where the latter serves as reference for the most localized states~\cite{Dodonov:2000aa}.} \sugg{The states that minimize $M_\infty$ are seen to be the most quantum ones. For a spin, when this distance is taken to be that of the closest mixture of coherent states to a given pure state, the most quantum states will be called the Queens of Quantumness.}

Coherent states clearly attain the maximal value
\begin{equation}
{M_{\infty}} ( |\beta \rangle ) = 1 \, .
\label{eq:Qmaxcoh}
\end{equation}
{For more general Gaussian states, we can restrict ourselves to the squeezed vacuum, since the displacement of states does not influence their values of $Q_{\mathrm{max}}$. We then have }
\begin{equation}
{M_{\infty}} ( | \xi \rangle ) = \text{sech} \, r 
\simeq \frac{1}{\sqrt{\bar{n}}} \, ,
\qquad \qquad \bar{n}=\sinh^2 r \gg 1 \, ,
\end{equation}
showcasing their high degrees of quantumness.

For number states, a direct calculation gives
\begin{equation}
{M_{\infty}} ( | {n} \rangle )= e^{-n}\frac{n^{n}}{n!} 
\simeq \frac{1}{\sqrt{2\pi n}},
\qquad \qquad  n \gg 1 .
\end{equation}
The upper limit value \sugg{of quantumness, corresponding to the lowest value of $M_\infty$,} is reached  when $n \gg 1$.

As before, we can look at the intermediate photon-added coherent states $| \beta, m \rangle$.  Since adding $m$ photons always scales the $Q$ function by $| \alpha |^{2m}$ and a normalization constant~\cite{Marian:2020aa}, one can readily compute
\begin{equation}
{M_{\infty}} ( | \beta, m \rangle ) =  
\frac{ [ 
\tfrac{1}{2} {|\beta|^2} ( 1 + \sqrt{1+ {4m}/{|\beta|^2}} )]^{2m}} 
{m! \; L_m ( -|\beta|^2 )}
\exp [ - \tfrac{1}{4} {|\beta|^2} (\sqrt{1+ {4m}/{|\beta|^2}} - 
1 )^2 ]\, .
\end{equation}
The Laguerre polynomials decrease with $|\beta|^{2}$ and decrease with $m$; this immediately allows us to identify Fock states, with $n=m$ and $|\beta|^2=0$, as having the highest quantumness.

As a final example, we return to the cat states (\ref{eq:catpm}). Now, for the even cat states, we find
\begin{equation}
{M_{\infty}} ( | \mathrm{cat} \rangle_{+} ) =  
\sugg{ \begin{cases}
\text{sech} (\beta^{2}) , & 0\leq \beta\leq 1\\
\frac{1}{2}+\frac{1}{2}\exp\left(-2\beta^2\right) , & \beta \gg 1
\end{cases} \, ,}
\end{equation}
{and one sees that the quantumness grows continuously as $\beta $ grows \sugg{but is eventually saturated}. This accords with the decoherence rate for superpositions of coherent states increasing with increased separation between the components of the superposition (in the phase space)~\cite{Brune:1996aa,Malbouisson:2003aa}. 
\sugg{
Similar conclusions can be found by numerically calculating $M_\infty$ for almost all of the Yurke-Stoler states~\cite{Yurke:1986aa} 
\begin{equation}
 |\text{cat}_\theta\rangle \propto |\beta\rangle +\cos\theta |-\beta\rangle \, ,
 \end{equation} 
 where for all $\theta\neq (2k+1)\pi$ the quantumness grows from $M_\infty=1$ (for $\beta=0$) and asymptotically reaches $M_\infty=\tfrac{1}{2}$ in the limit $\beta\to\infty$. The odd cat states, in contrast, already possessing some quantumness in the limit $\beta\to 0$ because their vacuum components vanish, have $M_\infty$ range monotonically from $\exp\left(-1\right)\simeq  0.37$ to $\tfrac{1}{2}$ as $\beta$ increases from $0$ to infinity, with quantumness \textit{decreasing} with increased $\beta$ \cite{Malbouisson:2003aa}.}} 

\subsection{Metrological power}
\label{sec:CV metrological power}

The field of quantum metrology, in the broadest sense, concerns itself with inventing useful nonclassical states and extracting from them some requisite metrological content. Loosely speaking, the metrological power of a quantum state is the metrological advantage that can be attributed to that state relative to classical limits~\cite{Tan:2019aa,Kwon:2019aa,Tan:2020aa}. 

The basic elements of parameter estimation can be found in a number of comprehensive reviews~\cite{lnp:2004uq,Paris:2009aa,Giovannetti:2011aa,Rafal:2012aa,Szczykulska:2016aa,Braun:2018aa,Sidhu:2020aa,Albarelli:2020aa}. The most fundamental element of quantum metrology is the probe state, which is represented by some density operator $\op{\varrho}$. We are often interested in estimating a single parameter $\lambda$ encoded via some quantum channel $\mathcal{E}_{\lambda}$. Information about $\lambda$ is extracted by passing a state $\op{\varrho}$ through the quantum channel $\mathcal{E}_{\lambda}$, resulting in the state transformation $\mathcal{E}_{\lambda} (\op{\varrho}) = \op{\varrho}_{\lambda}$. For simplicity, we will restrict our attention to unitary channels, so that $\op{\varrho}_{\lambda} = \op{U}_{\lambda} \, \op{\varrho} \op{U}_{\lambda}^{\dagger}$. The unitary operator $\op{U}_{\lambda}$ can be written as $\op{U}_{\lambda}= \exp (- i \lambda \op{G})$, where $\op{G}$ is a Hermitian operator that is called the generator of the unitary transformation.
 
The information about $\lambda$ is thereby encoded onto the probe $\op{\varrho}$. In order to estimate $\lambda$, we perform a measurement on $\op{\varrho}$, which is represented by some set of positive operator-valued measures (POVMs)~\cite{Holevo:1982aa} $\op{M} =\{ \op{\Pi}_{x} \}$. The latter comprise a set of positive semi-definite Hermitian operators that resolve the identity; that is,
\begin{equation}
\op{\Pi}_{x} \ge 0 \, , 
\qquad
\op{\Pi}^{\dagger}_{x} = \op{\Pi}_{x}^{\dagger} \, ,
\qquad
\int dx \; \op{\Pi}_{x} = \op{\openone} \, .
\end{equation}
By performing a measurement, we obtain a statistical distribution $P(x | \lambda) = \Tr (\op{\varrho}_{\lambda} \, \op{\Pi}_{x})$. Afterward, what remains is to obtain the best estimate of $\lambda$ given 
$P(x | \lambda)$, which is done using an estimator $\check{\lambda}$. Since we are trying to estimate the value of a fixed parameter $\lambda$, our guess should be correct on average if we repeat the experiment enough times.  This means that we should have
\begin{equation}
\langle \check{\lambda} \rangle = \int dx \; P(x|\lambda ) \, \check{\lambda}(x) = \lambda \, .
\end{equation}
When this happens, the estimator $\check{\lambda}$ is called unbiased.

There is an infinite number of possible POVMs that we can consider in quantum mechanics. It is therefore natural to ask whether there is an optimal measurement $\op{M}$ that we should perform on the state $\op{\varrho}_{\lambda}$. Helstrom, in his classical work~\cite{Helstrom:1976aa}, demonstrated that this question can actually be answered using fairly elementary arguments. To this end, we introduce the quantum Fisher information by~\cite{Petz:2011aa}
\begin{equation}
\Fish (\op{\varrho}, \lambda ) = \Tr (\op{\varrho}_{\lambda} \mathbb{D}^{2}_{\lambda}) \,,
\end{equation}
where $\mathbb{D}^{2}_{\lambda}$ is the so-called symmetric logarithmic derivative, defined implicitly via
\begin{equation}
\frac{\partial \op{\varrho}_{\lambda}}{\partial \lambda} = 
\tfrac{1}{2} \{\sugg{\op{\varrho}}, \mathbb{D} \} \, , 
\end{equation}
where $\{ \cdot, \cdot \}$ stands for the anticommutator $\{ \op{A}, \op{B} \} = \op{A} \op{B} + \op{B} \op{A}$. 

For unitary processes, with generator $\op{G}$, the quantum Fisher information simplifies to~\cite{Paris:2009aa}
\begin{equation}
\Fish (\op{\varrho}, \lambda ) = 2
\sum_{i,j} \frac{(p_{i} - p_{j})^{2}}{p_{i} + p_{j}} 
| \langle i |\op{G}| j \rangle |^{2} \, ,
\end{equation}
where $p_{i}$ and $| i \rangle$ are the eigenvalues and eigenvectors of $\op{\varrho}$, respectively. 

The quantum Fisher information $\Fish (\op{\varrho}, \lambda )$ is then optimized over all the measurements. As such, it depends only on the state. 
The crux of quantum metrology is the quantum version of the time-honoured Cram\'er-Rao lower bound~\cite{Rao:1945aa,Cramer:1946aa}: for any unbiased estimator $\check{\lambda}$, we have
\begin{equation}
\Var_{\op{\varrho}} (\check{\lambda}) \ge \frac{1}{\Fish (\op{\varrho}, \lambda)} 
\label{eq:qCRB}
\end{equation}
per single-detection event. This quantum Cramer-Rao bound may always be saturated, at least in principle. Recall that the symmetric logarithmic derivative $\mathbb{D}$ is a Hermitian matrix, which is diagonalizable. As such, we can consider its {spectral decomposition} $\mathbb{D} = \sum_{x} d_{\lambda,x} | \varphi_{\lambda,x} \rangle \langle \varphi_{\lambda,x} |$.  Since the bound can be saturated, this suggests that the quantum Fisher information  quantifies the usefulness of a probe $\op{\varrho}$ for the measurement of a given $\lambda$. 

For the case of a pure state, we have that
\begin{equation}
\Fish (\psi, \lambda ) = 4 \Var_{\psi} (\check{\lambda}) \, ,
\end{equation}
and, in consequence, the most quantum states, from this metrological point of view, are those with maximal quantum Fisher information. For unitary dynamics where the eigenvalues of $\op{G}$ are bounded, one may immediately identify the extremal quantum states: they are~\cite{Braunstein:1996aa}
\begin{equation}
|\psi_{\mathrm{opt}} \rangle = \frac{1}{\sqrt{2}} 
(| r_{\mathrm{max}} \rangle + e^{i \chi } | r_{\mathrm{min}})
\end{equation}
where $| r_{\mathrm{max}} \rangle$ and $| r_{\mathrm{min}} \rangle$ are the eigenvectors of $\op{\varrho}$ with maximal and minimal eigenvalues, respectively, and $\chi$ is an arbitrary phase.

Later on, we will be interested in estimating rotations. The natural counterpart in the case of a single mode is estimating a linear displacement. Let us consider the displacement 
\begin{equation}
\op{D} (r, \theta) = \exp[ - i r (e^{i \theta} \op{a} + 
e^{- i \theta} \op{a}^{\dagger})/\sqrt{2}] =
\exp (- i r \op{x}_{\theta} ) \, ,
\end{equation} 
where $\op{x}_{\theta} =  (e^{i \theta} \op{a} + 
e^{- i \theta} \op{a}^{\dagger})/\sqrt{2}$ represents a general rotated quadrature of the field. This has the general form of the displacement operator $\op{D} (\alpha)$ with $\alpha = r e^{i \theta}/\sqrt{2}$. Now, imagine we want to estimate the magnitude of the displacement $r$; this is equivalent to choosing $\op{x}_{\theta}$ as the generator of the transformation. Consequently, we have 
\begin{equation}
\Fish (\op{\varrho}, r, \theta) = 
 4
( \cos \theta \quad -\sin \theta)
\left ( 
\begin{array}{cc}
\Var_\psi (\op{x}) & \Cov_\psi (\op{x},\op{p}) \\
\Cov_\psi (\op{x},\op{p}) & \Var_\psi (\op{p})
\end{array}
\right )
\left ( 
\begin{array}{c}
\cos \theta \\
-\sin \theta 
\end{array}
\right )
, \label{eq:CV Fisher}
\end{equation} 
where $\Cov_\psi (\op{A},\op{B})= \tfrac{1}{2} \langle \{ \op{A}, \op{B} \} \rangle_\psi - \langle\op{A} \rangle_\psi \langle \op{B} \rangle_\psi$. 
Since the information about the phase $\theta$ is irrelevant, we can average over that variable; the result is
\begin{equation}
\bar{\Fish} (\op{\varrho}) =  \tfrac{1}{2}
[\Fish (\op{\varrho}, r, x) + \Fish (\op{\varrho}, r, p) ] \, ,
\end{equation} 
where $\Fish (\op{\varrho}, x)$ and $\Fish (\op{\varrho}, p)$  are the quantum Fisher informations for the corresponding quadratures. For pure states this becomes
\begin{equation}
\bar{\Fish} (\op{\varrho}) = 2 [ \Var_{\psi} ( \op{x}) +  \Var_{\psi} (\op{p}) ] 
= 2 + 4 [ \langle \op{a}^{\dagger} \op{a} \rangle_\psi - 
| \langle \op{a} \rangle_\psi |^2 ] \, .
\end{equation} 
This reduces to a number of other quantumness measures of a state, including the \sugg{total noise~\cite{Schumaker1986,Hillery1989}, total variance~\cite{Yadin:2018aa}, mean quadrature variance~\cite{Kwon:2019aa}, and} quadrature coherence scale~\cite{Hertz:2020aa}. 
For a coherent state $\bar{\Fish} (| \alpha \rangle ) = 2$ and for  any classical state the inequality $\bar{\Fish} ( \op{\varrho}_{\mathrm{cl}} ) \le  2$ holds. 

Clearly, the most quantum states are those maximizing $\bar{\Fish} ( \op{\varrho} ) $. For a given $\bar{n}= \langle \op{a}^\dagger\op{a} \rangle$, any state with $\langle\op{a} \rangle=0$ maximizes $\bar{\Fish}$; that is, the states with the greatest metrological power are those with 
\begin{equation}
   \int_\mathbb{C} d\alpha \, \alpha Q(\alpha) = 0 \,.
\end{equation} 
The first-order moments of these states vanish, which makes the latter treat all quadratures equally. In turn, higher-order moments of the states govern the metrological power through $\bar{n}$. This is a recurring theme that we will again see with the $\mathfrak{su}$(2) algebra. We see that coherent states have the least metrological power while states with vanishing first-order moments have the greatest metrological power.

There is more to the metrological-power story than averaging the quantum Fisher information over different experimental configurations. It is insightful to average the variances of all estimates of the displacement $r$ by averaging the quantum Cram\'er-Rao bound over all phases $\theta$:
\begin{equation}
\overline{\Var}_{\psi} (r) \geq \frac{1}{2\pi}\int_0^{2\pi}d\theta\,
\frac{1}{\Fish(\op{\varrho},r, \theta )} \,.
\end{equation}
Since variances are always positive, $1/\Fish(\op{\varrho},\op{x}_\theta)$ is a convex functional of $\Fish(\op{\varrho},r, \theta )$, allowing use of Jensen's inequality~\cite{Jensen:1906aa}
\begin{equation}
\frac{1}{2\pi}\int_0^{2\pi}d\theta\,\frac{1}{\Fish(\op{\varrho},r, \theta )} \geq 
\frac{1}{\bar{\Fish}(\op{\varrho})}=\frac{1}{2+4\bar{n}-4 |\langle \op{a} \rangle_\psi |^2} \,.
\label{eq:JensenCV}
\end{equation} 
Our convex functional is nonlinear; Jensen's inequality is then saturated if and only if   $\Fish(\op{\varrho},r, \theta )$ is independent of {the direction $\theta$.} This is only the case when the covariance matrix of $\op{x}$ and $\op{p}$ found in \eqref{eq:CV Fisher} is isotropic (i.e., proportional to the identity matrix). Isotropy of the covariance matrix is equivalent to $\Var_\psi(\op{a})=0$, which is satisfied by coherent states, number states, and more. 

Even though the second-order moments of both coherent states and number states are isotropic, the former have much less metrological power because their first-order moments are maximal. This may be seen by directly integrating \eqref{eq:JensenCV}, which can be formulated in a number of insightful ways 
\begin{equation}
\frac{1}{2\pi}\int_0^{2\pi}d\theta\,\frac{1}
{\Fish(\hat{\varrho}, r, \theta)}  = 
 \frac{1}{4\sqrt{\Var_\psi(\op{x}) \Var_\psi(\op{p})-
 \Cov_\psi(\op{x},\op{p})^2}} =
\frac{1}{2\sqrt{(1+2\bar{n}-2 |\langle \op{a} \rangle_\psi |^2 )^2-
4 |\Var_\psi(\op{a}) |^2}} \, .
\end{equation}
The Schr\"odinger-Robertson uncertainty relation ensures that~\cite{Dodonov:1987aa} 
\begin{equation}
\Var_\psi (\op{x} ) \, \Var_\psi(\op{p}) - 
\Cov_\psi(\op{x},\op{p})^2 \geq 
\frac{1}{4} \, ,
\end{equation} 
with equality if and only if $| \psi \rangle$ is an eigenstate of $\op{x}+\gamma\op{p}$ for some complex constant $\gamma$. The only possible normalizable states with this condition are \sugg{Gaussian states}; i.e., minimizing metrological power is equivalent to being a \sugg{squeezed} coherent state. We can also explicitly see that maximizing the metrological power is equivalent to setting the first-order moments $\langle \op{x} \rangle_\psi$ and $\langle \op{p} \rangle_\psi$ to zero and making the second-order moments isotropic: 
\begin{equation}
\Var_\psi(\op{x}) = \Var_\psi(\op{p}) \,, 
\qquad \qquad 
\Cov_\psi(\op{x},\op{p})= 0 \, .
\end{equation} 
This is in turn equivalent to the vanishing of the first- and second-order moments of $\op{a}$:
\begin{equation}
\int_\mathbb{C} d\alpha \, \alpha \; Q(\alpha) = 0 \,, 
\qquad \qquad 
\int_\mathbb{C} d\alpha \, \alpha^2 \; Q(\alpha) = 0 \, .
\end{equation}

From this it is clear that states with maximal first-order moment $\langle \op{a} \rangle_\psi$ (i.e., coherent states) have the least metrological power; whereas, states with vanishing first- and second-order moments of $\op{a}$ have the greatest metrological power. \sugg{Moreover, while a naive averaging of the Fisher information implies the large metrological power of squeezed states, an average over the Cram\'er-Rao bound shows that their metrological advantage for particular measurements is perfectly balanced by their disadvantage for other measurements, making them only as useful as coherent states.} 

\sugg{One way of identifying  states with the greatest metrological power} is from the symmetries of their Husimi functions. If the Husimi function is unchanged by a rotation in the complex plane $Q_\psi(\alpha)=Q_\psi(e^{-i\phi}\alpha)$, its moments must obey
\begin{equation}
\int_\mathbb{C} d\alpha \, \alpha^{k} \; Q_n(\alpha) = 
e^{ ik \phi} \int_\mathbb{C} \sugg{d\alpha} \; \alpha^{k}Q \; _n(\alpha)  \, .
\end{equation}
The $k$th moment must vanish unless \sugg{$e^{ik \phi }=1$}. The rotational symmetries given by $\phi$ determine the metrological power of a state.

Cat states \eqref{eq:catpm}, for example, have Husimi functions that are symmetric under $\alpha\to-\alpha$; they thus have vanishing first-order moments but nonzero $k=2$ moments. Similarly, the more sophisticated \sugg{four-component compass} state~\cite{Zurek:2001aa}  $\propto |{\alpha} \rangle + | -\alpha \rangle + 
 | i \alpha \rangle +  |-i\alpha \rangle$ has a rotational symmetry with $\phi=\pi/2$, and so all of its moments up until $k=4$ vanish.

Number states, at the far extreme, have all of their moments vanish 
\begin{equation}
\int_\mathbb{C} \sugg{d\alpha} \; \alpha^{k}Q_n(\alpha) 
=0
\end{equation}
due to the continuous polar symmetry $Q_n(\alpha)=Q_n ( |\alpha |)$ of their Husimi functions. A different kind of probe may be useful to showcase the advantages of states whose higher-order moments are also isotropic.

\section{Quantumness for spin variables}
\label{sec:spins}

{Quantum information can also be encoded in discrete degrees of freedom. The focus for this discrete-variable (DV) approach is on exploiting two-level systems---qubits---to generate a quantum analogue of the classical bit~\cite{Chuang:2000aa}. Common examples involve single photons living in a finite-dimensional space spanned, for instance, by orthogonal polarization modes, orbital angular momentum, or the two paths followed in an interferometer. From a mathematical perspective, these cases can be understood within the formalism of SU(2) and so we will talk here about spin variables.  In this section we present our quantumness indicators paralleling our CV treatment as much as possible, portraying the nuanced similarities and differences between these two worlds.}

\subsection{Bloch coherent states}

In this section we derive the spin or Bloch coherent states, in a fashion similar to what we did for the canonical coherent states. Our Hilbert space will be any finite-dimensional Hilbert space in which SU(2) acts irreducibly.

We consider a system whose dynamical group is SU(2).  The corresponding Lie algebra $\mathfrak{su} (2)$ is spanned by the operators $\{ \op{S}_{x}, \op{S}_{y}, \op{S}_{z} \}$ satisfying the commutation relation 
\begin{equation}
[\op{S}_{x}, \op{S}_{y} ] = i  \op{S}_{z} 
\end{equation} 
and cyclic permutations thereof.  The Casimir operator is $\op{\mathbf{S}}^{2} = \op{S}_{x}^{2} + \op{S}_{y}^{2}+ \op{S}_{z}^{2} = S (S+1) \openone$, so the eigenvalue $S$, which is a nonnegative integer or half integer, labels the irreducible representations (irreps).

We employ a fixed irrep of spin $S$, with a $(2S+1)$-dimensional carrier space $\mathcal{H}_{S}$ spanned by the standard angular momentum basis $\{ |S, m\rangle \mid m= -S, \ldots, S\}$, whose elements are simultaneous eigenstates of $\op{\mathbf{S}}^{2}$ and $ \op{S}_{z}$:
\begin{equation}
  \op{\mathbf{S}}^{2} |S, m \rangle = S (S+1) |S, m \rangle \, ,
  \qquad
  \op{S}_{z} |S, m\rangle = m |S, m \rangle \, .
\end{equation}
The raising and lowering operators $\op{S}_{\pm} = \op{S}_{x} \pm i \op{S}_{y}$  act on the basis through
\begin{eqnarray}
\op{S}_{+} |S,m \rangle & = & 
\sqrt{(S-m)(S+m+1)} |S, m+1 \rangle \, , \nonumber \\
& & \\
\op{S}_{-} |S,m \rangle & = & 
\sqrt{(S+m)(S - m+1)} |S, m-1 \rangle \, ; \nonumber
\end{eqnarray}
the highest-weight state is $|S,S\rangle$ and is annihilated by $\op{S}_{+}$. The isotropy subgroup (i.e., the largest subgroup that leaves the highest-weight state invariant) consists of all the elements of the form $\exp( i \chi \op{S}_{z})$, so it is isomorphic to U(1). The coset space is then SU(2)/U(1), which is simply the unit sphere $\mathcal{S}_{2}$ (the so-called Bloch sphere); this is the classical phase space, which is the natural arena to describe the {physics}. 

{An element} of the quotient space SU(2)/U(1) can be represented as
\begin{equation}
\op{D} (\theta, \phi) = \exp (i \theta \op{S}_{z} ) \, 
\exp (i \phi \op{S}_{y} ) = \exp \left [ \tfrac{1}{2} \theta
  (\op{S}_{+}e^{-i\phi}- \op{S}_{-}e^{i\phi} ) \right ] \, ,
\end{equation}
which acts as a displacement operator on the sphere $\mathcal{S}_{2}$.  Note that $\op{D}^{\dagger} (\theta, \phi) = \op{D} (- \theta, \phi)$. If we denote by $\mathbf{n}$ the unit vector in the direction $(\theta, \phi)$; i.e., 
$\mathbf{n} = (\sin \theta \cos \phi, \sin \theta \sin \phi, \cos \theta)^{t}$, we define the SU(2) coherent states by~\cite{Arecchi:1972zr}
\begin{equation}
| \mathbf{n} \rangle = \op{D} (\mathbf{n}) |S, S \rangle \, .
\end{equation}
The displacement operator can alternatively be written in its normal disentangled form
\begin{equation}
\label{eq:disent}
\op{D} (\mathbf{n}) = \exp (\zeta \op{S}_{+}) \exp (\eta \op{S}_{z}) 
\exp(- \zeta^{\ast} \op{S}_{-}) \, ,
\end{equation}
with 
\begin{equation}
\label{eq:stereo}
\zeta = \tan (\theta/2) e^{i \phi} \, , 
\qquad
\eta = \ln ( 1 + | \zeta |^{2} ) \, .
\end{equation}
The relation between $\zeta$ and $\mathbf{n}$ is precisely a stereographic projection.  {In our definition of $| \mathbf{n} \rangle$, we take this projection to be from the north pole to the complex plane, though the opposite choice can often be found in the literature~\cite{Perelomov:1986ly,Gazeau:2009aa}. Using \eqref{eq:disent} we obtain}
\begin{equation}
|\mathbf{n} \rangle = \frac{1}{( 1 + | \zeta |^{2})^{S}} 
\exp (\zeta \op{S}_{+}) |S, S \rangle \, .
\end{equation}
On expanding the exponential, we {can write the coherent states as an expansion over the basis states of the irrep}:
\begin{equation}
|\mathbf{n} \rangle = \frac{1}{( 1 + | \zeta |^{2})^{S}} \sum_{m=-S}^{S} 
c_m \zeta^{S+m} |S, m\rangle \, ,
\end{equation} 
or, {employing} the stereographic projection (\ref{eq:stereo}), 
\begin{equation}
|\mathbf{n} \rangle =  \sum_{m=-S}^{S} 
c_{m} \; [\sin (\theta/2) ]^{S+m}  
[\cos (\theta/2) ]^{S-m} e^{-i (S+m) \phi}|S, m\rangle \, ,
\end{equation} 
with 
\begin{equation}
\sugg{c_{m} = \sqrt{\binom{2S}{S+m}} \, .}
\end{equation}
Note that the spin coherent state $|\mathbf{n} \rangle$ is an eigenvector of the operator $\op{S}_{\mathbf{n}} = \op{\mathbf{S}}\cdot \mathbf{n}$; i.e.,
\begin{equation}
\op{S}_{\mathbf{n}} |\mathbf{n} \rangle = -S |\mathbf{n} \rangle \, .
\end{equation} 

The system of spin coherent states is complete, but the states are not mutually orthogonal:
\begin{equation}
{| \langle \mathbf{n}_{1} | \mathbf{n}_{2} \rangle |^{2} =
[ \tfrac{1}{2} ( 1 + \mathbf{n}_{1} \cdot \mathbf{n}_{2} ) ]^{2S} \, .}
\end{equation} 
Still, they permit a resolution of the unity in the form
\begin{equation}
\frac{2S+1}{4 \pi} \int_{\mathcal{S}_{2}} d\mathbf{n} \; 
|\mathbf{n} \, \rangle \langle \mathbf{n} | = \openone \, ,
\end{equation}
with the {rotationally invariant measure given by $d\mathbf{n} = \sin\theta \, d\theta d\phi$.} With this completeness relation one is able to decompose an arbitrary state over the coherent states
\begin{equation}
| \psi \rangle = \frac{2S+1}{4 \pi} \int_{\mathcal{S}_{2}} d\mathbf{n} \;
{\psi (\mathbf{n}^{\ast})} \; |\mathbf{n} \rangle \, .
\end{equation}
Introducing the Dicke basis $\{ |S, m \rangle \}$, we obtain the wave function in the coherent-state representation {$\psi (\mathbf{n}^{\ast}) = \langle \mathbf{n} | \psi \rangle$ as}
\begin{equation}
{\psi( \mathbf{n}^{\ast})} = \frac{1}{( 1 + | \zeta |^{2})^{S}} 
\sum_{m=-S}^{S} c_{m} \psi_{m} \; \zeta^{s+m} \, . 
\end{equation}

The coherent states minimize the fluctuations of the Casimir operator
\begin{equation}
\Delta^{2} = \Var_{\mathbf{n}} (\op{S}_{x}) + 
\Var_{\mathbf{n}} (\op{S}_{y}) {+}
\Var_{\mathbf{n}} (\op{S}_{z}) =
\langle \op{\mathbf{S}}^{2} \rangle - \sum_{k=x, y, z} \langle \op{S}_{k} \rangle^{2} \,.
\label{eq:SU(2) total variance}
\end{equation}
{It is precisely in  this sense that they can be considered the closest to classical states.} 

\subsection{Husimi function and stellar representation}

We will define the Husimi $Q$-function for SU(2) in much the same way as we did for continuous variables; that is,
\begin{equation}
Q_{\varrho} (\mathbf{n}) = 
\langle \mathbf{n} | \op{\varrho} | \mathbf{n} \rangle \, ,
\end{equation}
and, for the particular case of a pure states, which will be our main interest in what follows, 
\begin{equation}
Q_{\psi} (\mathbf{n}) = 
| \langle \mathbf{n} | \psi \rangle |^{2}  \, .
\end{equation}
It is clear that $Q$ is positive and is normalized through 
\begin{equation}
\frac{2S+1}{4\pi} \int_{\mathcal{S}_{2}} d\mathbf{n} \; Q_{\varrho} (\mathbf{n}) = 1 \, ,
\end{equation} 
so it provides a genuine probability distribution on the sphere. $Q$ is bounded from above. Its maximum value has an interesting interpretation: the Fubini–Study distance between {$|\psi \rangle$ and $| \mathbf{n} \rangle$} is given by $\mathbb{D}_{\mathrm{FS}} = \arccos \sqrt{\kappa}$, where $\kappa = | \langle \mathbf{n} | \psi \rangle |^{2} = Q_{\psi} (\mathbf{n})$, so the maximum of $Q_{\psi} ( \mathbf{n})$ determines the minimum distance between $| \psi \rangle$ and the orbit of coherent states, {as we have already discussed in Sec.~\ref{sec:maxQ} in the context of CV.}  

The coherent-state wave function $\psi (\mathbf{n}^{\ast})$, also called the Bargmann wave function, is now a polynomial of degree $2S$, so it is uniquely characterized by its zeros: they constitute the Majorana constellation. 
{Several decades after its conception, this representation has recently attracted a great deal of attention in several fields~\cite{Hannay:1998aa,Hannay:1998ab,Ribeiro:2007aa,Makela:2010aa,Lamacraft:2010aa,Bruno:2012aa,Lian:2012aa,Devi:2012aa,Cui:2013aa,Yang:2015aa,Bjork:2015aa,Bjork:2015ab,Liu:2016aa,Chryssomalakos:2017aa,Chryssomalakos:2018aa,Goldberg:2018aa}.}
In consequence, we can factorize the associated polynomial and write the wave function as 
\begin{equation}
{\psi (\mathbf{n}^{\ast})} = \frac{Z^{\ast}_{2S}}{(1 + | \zeta |^{2})^{S}} 
(\zeta - \omega_{1}) \ldots (\zeta - \omega_{2S})\, ,
\end{equation}
where $Z^{\ast}_{2S}$ is the final component of the vector of coefficients of $|\psi\rangle$ in the basis of the irrep. Therefore, the Husimi $Q$ function reads
\begin{equation}
Q_{\psi} (\mathbf{n}) = \frac{|Z_{2S}|^{2}}{(1 + | \zeta |^{2})^{S}}
|\zeta - \omega_{1}|^{2} \ldots |\zeta - \omega_{2S}|^{2} \,. 
\end{equation}

Let us examine some relevant examples. For a coherent state {$|\mathbf{n}_{0} \rangle$}, we have
\begin{equation}
Q_{\mathbf{n}_{0}} (\mathbf{n}) = [\tfrac{1}{2} 
(1 + \mathbf{n} \cdot\mathbf{n_{0}})]^{2S} \, ,
\end{equation} 
so there is a single zero at $\mathbf{n} = - \mathbf{n}_{0}$ with multiplicity $2S$. Coherent states are the only ones for which the constellation collapses to a single point; i.e., they have the most localized constellation.
For the Dicke state $|S, m\rangle$, a simple calculation shows that
\begin{equation}
Q_{S,m} (\mathbf{n}) = \binom{2S}{S+m} 
[\sin (\theta/2) ]^{2(S+m)}   [\cos (\theta/2) ]^{2(S-m)} \, .
\end{equation}
For $m=S$ we have a coherent state and the zeros are at the north pole. For $m=0$ the function is concentrated as a ``{napkin ring}'' along the equator. The idea is that the Husimi function tends to be more spread out the more the state differs from being a coherent one. In other words, the most classical Husimi function seems to be the most localized one; whereas, the most quantum distribution should have the most spread constellation. 

Another convenient way of rewriting the Husimi function is as
\begin{equation}
\label{eq:prodpar}
Q (\mathbf{n}) = k_{S} \; \sigma (\zeta , \omega_{1}) \cdots
\sigma(\zeta , \omega_{2S}) \, ,
\end{equation}
where 
\begin{equation}
\sigma (\zeta , \omega ) = 
\frac{| \zeta - \omega |^{2}}{(1 + |\zeta |^{2})(1 + | \omega |^{2})} = 
\tfrac{1}{2} (1 - \cos {d} ) = \sin^{2} ({d}/{2}) = 
\tfrac{1}{4} d_{\mathrm{ch}}^{2} \, .
\end{equation}
Here  $d$ is the geodesic and $d_{\mathrm{ch}}$ the chordal distance between the points $\zeta$ and $\omega$ on the unit sphere~\cite{Baecklund:2014ng}. The factor $k_{S}$ is a $\zeta$-independent normalizing factor that can be expressed as the sum of a set of symmetric functions of the chordal distances~\cite{Lee:1988aa}.

\subsection{The multipolar expansion}
\label{sec:multipoles}

Instead of directly using the states $\{ | S, m \rangle \}$, it is more convenient to expand $\op{\varrho}$  as
\begin{equation}
  \label{rho1}
  \op{\varrho} =  \sum_{K= 0}^{2S} \sum_{q=-K}^{K}  
  \varrho_{Kq} \,   \op{T}_{Kq} \, ,
\end{equation}
where the irreducible tensor operators $T_{Kq}^{(S)}$ are~\cite{Fano:1959ly,Blum:1981ya,Varshalovich:1988ct}
\begin{equation}
  \label{Tensor} 
  \op{T}_{Kq} = \sqrt{\frac{2 K +1}{2 S +1}} 
  \sum_{m,  m^{\prime}= -S}^{S} C_{Sm, Kq}^{Sm^{\prime}} \, 
  |  S , m^\prime \rangle \langle S, m | \, ,
\end{equation}
with $ C_{Sm, Kq}^{Sm^{\prime}}$ being the Clebsch-Gordan coefficients
that couple a spin $S$ and a spin $K$ \mbox{($0 \le K \le 2S$)} to a
total spin $S$.  These tensors comprise an orthonormal basis {$
\Tr  (\op{T}_{K q}\, \op{T}_{K^{\prime} q^{\prime}}^{\dagger} )  =
 \delta_{K  K^{\prime}} \delta_{q q^{\prime}}$}
and have the correct transformation properties: under a rotation
parametrized by the Euler angles $(\alpha, \beta, \gamma)$, we have
\begin{equation}
  \op{R} (\alpha , \beta, \gamma) \, \op{T}_{Kq}  \,
  \op{R}^{\dagger} (\alpha , \beta, \gamma) =
  \sum_{q^{\prime}}  D_{q^{\prime} q}^{S} (\alpha, \beta, \gamma ) \,
  \op{T}_{Kq^{\prime}}  \, ,
  \label{eq:sym3}
\end{equation}
where the $D_{q^{\prime} q}^{S} (\alpha, \beta, \gamma) $ stands for {the Wigner $D$ matrix; that is, the matrix elements of the rotation operator $\op{R} (\alpha, \beta, \gamma)$ in the basis $ |S, m \rangle$~\cite{Varshalovich:1988ct}.}

Although at first sight these tensors might look a bit intimidating, they are nothing but the multipoles used in atomic physics~\cite{Blum:1981ya}.  After some calculations, one can recast 
Eq.~\eqref{Tensor} as
\begin{equation}
  \label{eq:multi}
  \begin{array}{l}
    \op{T}_{00} = \displaystyle     
    \frac{1}{\sqrt{2 S + 1}} \op{\openone} \, , \\
    \\
    \op{T}_{10} =   
    \displaystyle{\sqrt{\frac{3}{( 2 S + 1 ) (S+1) S}}} \; 
    \op{S}_{z} \, , 
    \qquad  
    \op{T}_{1\mp 1} = \displaystyle  
    \sqrt{\frac{3}{( 2 S + 1 ) (S+1) S}}  \; 
    \op{S}_{\pm} \, , \\
    \\
    \op{T}_{20} = \displaystyle{\sqrt{\tfrac{C}{6}}}  
    (3\op{S}_{z}^{2} - \op{S}^{2}) \, , 
    \qquad
    \op{T}_{2\mp 1}  =   \displaystyle{\sqrt{\tfrac{C}{2}}} \; 
    \{ \op{S}_{z},  \op{S}_{\pm} \}  \, , 
    \qquad
    \op{T}_{2\mp 2}  =   \displaystyle{\sqrt{\tfrac{C}{2}}}   \; 
    \op{S}_{\pm}^{2} \, ,
  \end{array}
\end{equation}
where $C=30/[(2S + 3)(2 S + 1) (2S-1) (S+1)]$.  In consequence, we conclude that $\op{T}_{Kq}$ can be related to the $K$th powers of the generators. The dipole $\varrho_{1q}$ is the first-order moment of $\op{\mathbf{S}}$ and thus corresponds to the classical picture, in which the state is represented by its average value on the Bloch sphere.  Complete characterization of a state demands the knowledge of all the multipoles. 

The corresponding expansion coefficients {$\varrho_{Kq} =  \Tr ( \op{\varrho} \, \op{T}_{Kq}^{\dagger} )$ are known as state multipoles. The hermiticity and positive semidefiniteness of $\op{\varrho}$ force the conditions}
\begin{equation}
 \varrho_{K-q} = ( -1)^{q} \, \varrho_{Kq}^\ast \, , \qquad  \qquad
  \sum_{q=- K}^{K} |\varrho_{Kq}^{(S)}|^{2} \leq  \Lambda_{K} \, ,
  \label{eq:cons}
\end{equation}
for every $K>1$ and $\Lambda_{K}$ a positive constant.  

Finally, we turn to the important class of axially-symmetric states~\cite{Blum:1981ya}. They are invariant under rotations about an axis that we take as the $z$ axis. Since $ D_{qq^{\prime}}^{S} 
(0,0,\gamma) = \exp(-i q \gamma) \delta_{qq^{\prime}}$, this implies
\begin{equation}
  \op{\varrho}_{\mathrm{axial}} = 
  \sum_{K=0}^{2S} \varrho_{K0} \; \op{T}_{K0}   \, .
  \label{eq:axsys}
\end{equation}
Thus, axially-symmetric states are characterized exclusively by the multipole components $\varrho_{K0}$.  Any density operator that can be obtained from $\op{\varrho}_{\mathrm{axial}} $ via an SU(2) transformation also represents an axially-symmetric state, as a rotation only changes the direction of the symmetry axis of the state.
Some axially symmetric systems are also invariant under reversal of the symmetry axis (i.e., $z \rightarrow -z$). As this corresponds to a rotation around the $y$ axis by an angle $\pi$ and $ D_{qq^{\prime}}^{S} (0, \pi , 0) = (-1)^{K+q} \delta_{q \, -  q^{\prime}}$, we find  $\varrho_{K0} = (-1)^{K} \varrho_{K0}$ from \eqref{eq:sym3}, so only multipoles of even rank $K$ contribute.

The Husimi $Q$ function encompasses complete information about a state, which is tantamount to knowing all the multipoles. This can be stressed if we rewrite $Q_{\varrho} (\mathbf{n})$ as~\cite{Klimov:2002cr}
\begin{equation}
  \label{eq:QsumK}
  Q_{\varrho} ( \mathbf{n} ) =  \sum_{K=0}^{2S} 
  Q_{\varrho}^{(K)} (\mathbf{n} ) \, ,
\end{equation}
where 
\begin{equation}
  \label{eq:QSU2rj}
  Q_{\mathbf{\varrho}}^{(K)} (\mathbf{n} ) =  \frac{\sqrt{4 \pi}}{\sqrt{2S+1}} 
  \sum_{q=-K}^{K}  C_{SS,K0}^{SS} \, \varrho_{Kq} \,
  Y_{Kq}^{\ast} ( \mathbf{n} ) \, .
\end{equation}
{Here,} $ Y_{Kq}^{\ast} ( \mathbf{n} )$ are the spherical harmonics  
and the Clebsch-Gordan coefficient $C_{SS,K0}^{SS}$ has a closed analytical form~\cite{Varshalovich:1988ct}
\begin{equation}
  \label{eq:Cesp}
  C_{SS,K0}^{SS} = \frac{\sqrt{2S+1} (2S)!}
  {\sqrt{(2S-K)! \, (2S+1 + K)!}} \, .
\end{equation}
In this way, the $Q$ function of the state appears as a sum  of partial components {$Q_{\varrho}^{(K)}$, which inherit} the properties of $Q_{\varrho}$, but the partial components exclusively contain information about the $K$th  moments of the Stokes variables. Equation~(\ref{eq:QsumK}) appears as an appropriate tool to arrange the successive moments.

We illustrate this viewpoint with the simple example of the state represented by $|1, 0 \rangle$ in the angular momentum basis. As discussed before, its $Q$ function can be immediately calculated 
\begin{equation}
  \label{eq:Qt11}
  Q_{|1, 0 \rangle} ( \mathbf{n} ) = \sin^{2} \theta  \, .
\end{equation}
It is independent of $\phi$ and its shape is an equatorial {napkin ring}, revealing that the state is highly delocalized on the Bloch sphere.  The partial components are
\begin{equation}
  \label{eq:Qp11}
  Q_{|1, 0 \rangle}^{(0)} ( \mathbf{n} ) = \frac{1}{3} \, ,
  \qquad 
  Q_{|1, 0 \rangle}^{(1)} ( \mathbf{n} ) = 0 \, ,
  \qquad 
  Q_{|1, 0 \rangle}^{(2)} ( \mathbf{n} ) =  \frac{2}{3} -  \cos^{2} \theta  
  \, .
\end{equation}
The sum of these three terms gives, of course, the result Eq.~(\ref{eq:Qt11}), but there is more information encoded in Eq.~(\ref{eq:Qp11}): the dipolar contribution is absent, confirming that this state conveys no first-order information. This is the reason why this was the first state in which hidden polarization was detected~\cite{Usachev:2001ve}. 

To conclude, we note that, since the spherical harmonics are a complete set of orthonormal functions on $\mathcal{S}_2$, they may be  used to expand the Husimi function $Q(\bf{n})$; the resulting coefficients turn out to be the multipoles
\begin{equation}
  \varrho_{Kq}=\mathcal{C}_K \int_{\mathcal{S}_2} 
 d\mathbf{n}\; {Y}_{Kq}(\mathbf{n}) \; Q(\mathbf{n}),
  \end{equation}
where the normalization constant is 
\begin{equation}
  \mathcal{C}_K=\sqrt{\frac{4\pi}{2S+1}}\frac{1}{C_{SS,K0}^{SS}} \, , 
  \end{equation}
to agree with previous definitions. 
When expressed in the Cartesian basis these multipoles appear in a
very transparent way. For example, the three dipole $(\varrho_{1q})$
and the five quadrupole $(\varrho_{2q})$ terms can be given,
respectively, by:
\begin{equation}
  \wp_i=\langle n_i \rangle, \qquad  \qquad 
\mathcal{Q}_{ij}=\langle 3 n_i n_j - \delta_{ij}\rangle,
  \end{equation} 
where {the expectation values of a function $f( \mathbf{n})$ of the vector $\mathbf{n}$ are calculated with respect to the Husimi $Q$ function; i.e.,}
\begin{equation}
 \langle f(\mathbf{n})\rangle= 
\frac{\int_{\mathcal{S}_2}\; d\mathbf{n} \;
f (\mathbf{n})  Q(\mathbf{n})}
{\int_{\mathcal{S}_2}\;d\mathbf{n}\;Q(\mathbf{n})} \, .
\end{equation}
Therefore, the state multipoles appear as the standard ones in
electrostatics, replacing the charge density by $Q(\mathbf{n})$
and distances by directions~\cite{Jackson:1999aa}. They are the $K$th-directional moments of the state and, therefore, these terms resolve
progressively finer angular features.

\subsection{Wehrl entropy}

Following our approach in the previous section, we can go on to define 
the Wehrl entropy for an SU(2) state as
\begin{equation}
S_{\mathrm{W}} (\op{\varrho}) = - \frac{2S+1}{4 \pi}
\int_{\mathcal{S}_{2}} d\mathbf{n} \; Q_{\varrho} (\mathbf{n})
\; \ln Q_{\varrho} (\mathbf{n}) \, .
\end{equation} 
As we have seen in the continuous-variable case, one of the key properties of the Husimi $Q$-function, as proved by Lieb, is that the Wehrl entropy attains its minimum for the coherent states. Clearly, a natural question would be whether the same is true for the $Q$-function on the sphere. Indeed, Lieb conjectured that 
\begin{equation}
S_{\mathrm{W}} (\op{\varrho}) \geq \frac{2S}{2S+1} \, ,
\end{equation} 
and that the equality holds if and only if the state is a coherent state. {This remained an open problem for 35 years, until the very recent demonstration of its veracity by Lieb and \sugg{Solovej}~\cite{Lieb:2014aa}.} 

The parametrization \eqref{eq:prodpar} is especially germane for this topic, as the logarithm turns into a sum of symmetric functions:
\begin{equation}
\label{eq:alperm}
 S_{\mathrm{W}} (\op{\varrho}) = - \frac{2S+1}{4 \pi}
\int_{\mathcal{S}_{2}} d\mathbf{n} \; Q_{\varrho} (\mathbf{n})
\; \left [ \ln k_{S} + \sum_{i=1}^{2S} \ln \sigma (\zeta, \omega_{i} ) \right ] \, . 
\end{equation}
In the particular case of a coherent state, the $2S$ roots are identical, which implies that $\sigma (\omega_{i}, \omega_{j}) = 0$ for any pair of unit vectors, whereupon the Wehrl entropy equals $ S_{\mathrm{W}} = 2S/(2S + 1)$, in agreement with the Lieb conjecture.  

In general, \eqref{eq:alperm} can be formally solved in terms of various symmetric functions of the squares of the chordal distances. However, writing down all the symmetric functions that occurs for high values of $S$ is a formidable task.

The complementary problem of states that maximize the Wehrl entropy is accordingly an open problem.  It is clear that the Wehrl entropy increases as the points on the sphere spread further apart and attains a maximum for the most quantum states; i.e., when the points on the unit sphere are as far from one another as possible.  The basic problem of maximizing the Wehrl entropy then consists of distributing points on a sphere in order to optimize some function (the Wehrl entropy) that depends on the positions of the points. The distribution of points that corresponds to an optimized value of the function may not be unique, even though the criterion that the points are as far as possible from one another remains. Numerical results for some specific values of the dimension $S$ can be found in Ref.~\onlinecite{Baecklund:2014ng}; we present some of them in Fig. \ref{fig:extstat}.

\subsection{Inverse participation ratio}

The linear Wehrl entropy in this case is a literal translation of \eqref{eq:SWlinCV} to the sphere, namely 
\begin{equation}
S_{\mathrm{W, lin}} (\op{\varrho}) = 1 - \frac{2S+1}{4 \pi}
\int_{\mathcal{S}_{2}} d\mathbf{n} \; Q^{2} (\mathbf{n} ) ,
\end{equation} 
and much in the same way the second moment 
\begin{equation}
M_{2} (\op{\varrho}) = \frac{2S+1}{4 \pi} 
\int_{\mathcal{S}_{2}} d\mathbf{n} \; Q^{2} (\mathbf{n}) \, ,
\end{equation}
whose inverse is the inverse participation ratio. 

In the case of the sphere, the expansion \eqref{eq:QsumK} proves to be especially appropriate. In fact, using that formula,  {$M_{2} (\op{\varrho} ) $ can be expressed as}
\begin{equation}
M_{2} (\op{\varrho}) = \sum_{K=0}^{2S} \sum_{q=-S}^{S}
\left ( C_{SS,K0}^{SS}\right )^{2} |\varrho_{Kq}|^{2} \, .
\end{equation} 
If the Clebsch-Gordan coefficients $C_{SS,K0}^{SS}$ were absent from this expression, $M_{2} (\op{\varrho})$ would reduce to the purity (or, equivalently, to the integral of the square of the Wigner function $W$ over the sphere).  In other words, it is the Clebsch-Gordan coefficients that convey the quantum properties to this problem.

Following the same procedure as in Ref.~\onlinecite{Bjork:2015aa}, one can show that  this quantity is optimized by SU(2) coherent states, confirming that they are the most localized states. 

\subsection{Husimi extrema}

{The definition from \eqref{eq:geometric quantum Marian} applied to our case yields for a pure state}
\begin{equation}
{M_{\infty} ( | \psi  \rangle ) = \max_{\mathbf{n}} Q_\psi (\mathbf{n}) \, . }
\label{eq:M inf}
\end{equation} 
{This can be understood as the Hilbert-Schmidt distance between a state and the set of pure classical states. By construction, coherent states minimize this distance, so they are the least quantum states.}

{The most quantum states according to this measure have been found in the context of the geometric measure of entanglement~\cite{Hubener:2009aa,Aulbach:2010jw,Baguette:2014ws}. Finding states that minimize the maximum in \eqref{eq:M inf} is achieved by restricting the search to pure  states  with specific rotational symmetries or to directions $\mathbf{n}$ pointing in a subset of all directions, as these restrictions are guaranteed to contain the optimal results~\cite{Aulbach:2004aa}. We present the results for some dimensions in  Fig.~\ref{fig:extstat}.}

{An alternative criterion is the Hilbert-Schmidt distance between a state and the set of classical states, the latter of which can be regarded as the convex hull of those density matrices that can be decomposed as positive-weight sums of projectors onto coherent states. 
Similarly to $M_\infty$, coherent states minimize this measure, so they are the least quantum states. The resulting most quantum states go by the name of the \emph{Queens of Quantumness} and have been carefu\sugg{l}ly studied in Ref.~\onlinecite{Giraud:2010db}; these are different from the previous states because there are pure states whose closest classical state is mixed. For a fixed state, we represent the matrix  minimizing the distance as a linear combination of coherent states whose directions densely and uniformly cover the unit sphere, $\varrho_c = \sum_{i=1}^{N} \nu_{i} |\mathbf{n}_{i} \rangle \langle \mathbf{n}_{i}|$  with large $N$. Then $M_{\infty}$ becomes a quadratic function of the coefficients $\nu_{i}$, which has to be minimized under the constraints $\nu_{i} \ge 0$ and $\sum_{i} \nu_{i} =1$. This is a quadratic program that can be solved by a variety of algorithms, and the solutions seem to be unique. We compare the results for some dimensions in Fig.~\ref{fig:extstat}.}

\subsection{Cumulative multipolar distribution}

The quantity $\sum_{q} | \varrho_{Kq} |^{2} $ gauges the overlap of a state with the $K$th multipole pattern.  For most states, only a limited number of multipoles play a substantive role and the rest of them have an exceedingly small contribution. Therefore, it seems more convenient to look at the cumulative distribution~\cite{Hoz:2013om}
\begin{equation}
\mathcal{A}_{M} = 
\sum_{K= 1}^{M} \sum_{q=- K}^{K} | \varrho_{Kq} |^{2} \, ,
\end{equation}
which concisely condenses the state's angular properties up to order $M$ ($1 \le M \le 2S$). Observe that we omit the monopole, as it is a constant term, and that when $M$ goes to $2S$, $\mathcal{A}_{2S}$ is just the purity of the state. 

The distribution $\mathcal{A}_{M}$ can be regarded as a nonlinear functional of the density matrix $\op{\varrho}$. On that account, one can try to ascertain the states that maximize $\mathcal{A}_{M}$  for each order $M$.  We shall be considering only pure states, with coefficients $\Psi_{m}$ in the basis $ | S, m \rangle$.  We easily find 
\begin{equation}
\label{eq:AMS}
\mathcal{A}_{M} =\sum_{K=1}^{M} \sum_{q=-K}^{K}
\frac{2K+1}{2S+1} 
\left |  \sum_{m,m^{\prime }=-S}^{S} C_{Sm,Kq}^{Sm^{\prime} }
\Psi_{m^{\prime}} \Psi_{m}^{\ast } \right |^{2} \, .
\end{equation}
It has been shown that the maximum value is 
\begin{equation}
  \mathcal{A}_{M} =  \frac{2S}{2S +1} -
  \frac{[\Gamma (2S + 1)]^{2}}{\Gamma (2S-M) \Gamma (2S + M +2)} \, ,
\end{equation}
and this happens for  SU(2) coherent states for all orders $M$.

Next, we concentrate on minimizing $\mathcal{A}_{M}$. Obviously, the maximally mixed state  $\op{\varrho} = \textstyle{\frac{1}{2S+1}} \op{\openone}_{2S+1}$ zeroes all of the multipoles and so indeed causes $\mathcal{A}_{M}$ to vanish for all $M$, being fully unpolarized~\cite{Prakash:1971fr,Agarwal:1971zr}.  Nonetheless, we are interested in pure $M$th-order unpolarized states.  The strategy we adopt is thus very simple to state: starting from a set of unknown normalized state amplitudes in Eq.~\eqref{eq:AMS}, which we write as $\psi_{m} = a_{m}+i b_{m}$ ($a_{m}, b_{m} \in \mathbb{R}$), we try to make $\mathcal{A}_{M}=0$ for the highest possible $M$.  This yields a system of polynomial equations of degree two for $a_{m}$ and $b_{m}$, which we solve using Gr{\"o}bner bases implemented in the computer algebra system {\sc magma}~\cite{Bosma:1997xp}.  In this way, we find exact algebraic expressions and we can detect when there is no feasible solution.

The resulting states (which, in some cases, are not unique)  have been termed the \emph{Kings of Quantumness} and a complete list can be found at \url{http://polarization.markus-grassl.de}, with their associated Majorana constellations.  

Intuitively, one expects that these constellations have their points arranged as symmetrically as possible on the unit sphere, and this is the case.
These states, initially dubbed anticoherent states~\cite{Zimba:2006fk} are in a sense the opposite of SU(2) coherent states: while the latter correspond as nearly as possible to a classical spin vector pointing in a given direction, the former \emph{point nowhere}; i.e., the average spin vector vanishes and the fluctuations up to order $M$ are isotropic.  

\begin{figure}
    \centering
    \includegraphics[width=0.90\textwidth]{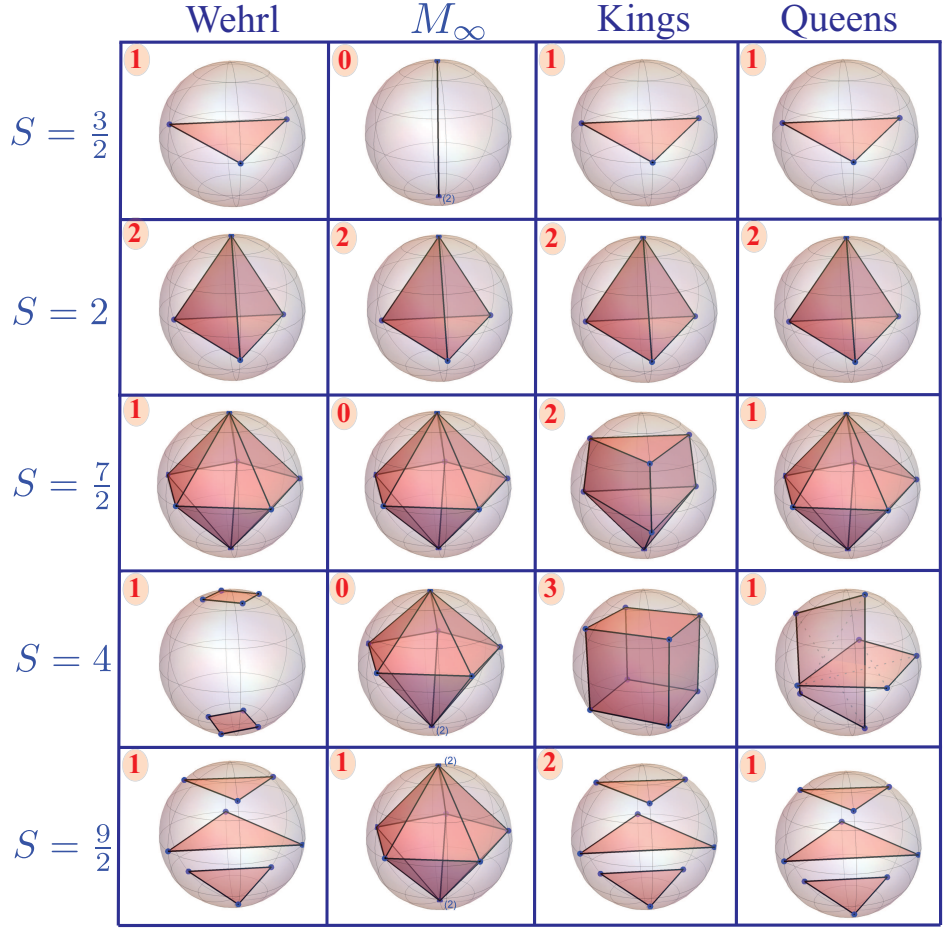}
    \caption{Majorana representation for states that maximize the Wehrl entropy, maximize $M_{\infty}$, are Kings of Quantumness, and are Queens of Quantumness. {In each row, we indicate the value of $S$. In the top-left corner of each cell we indicate (in red) the corresponding degree of unpolarization. The degeneracies at some points of the constellations are indicated in blue.}}
    \label{fig:extstat}
\end{figure}

{The configurations and the degrees of unpolarization of the corresponding states are also shown in Fig.~\ref{fig:extstat}. We see that for small $S$ the configurations are almost identical  for all the extremal principles. For larger $S$, they differ in general and the degree of unpolarization is also markedly different.}
\sugg{This highlights the need for a \emph{Royal Family} of states, where each member is the most quantum according to a unique principle that distinguishes it from the agreed-upon least classical states. The Kings and Queens are but examples of the necessity for specifying a principle before discussing quantumness.}

\subsection{{Other spherical configuration problems}}
\label{Other problems}

As we have seen, the extremal principles thus far can be reinterpreted as distributing $N$ points on a sphere in the \emph{most symmetric} way. This problem has a long history and many different solutions, depending on the cost function one tries to optimize~\cite{Conway:1996ys,Saff:1997gj}. To complete our previous discussion, we shall focus on a few related formulations.

The first problem uses the concept of spherical $t$-designs~\cite{Delsarte:1977dn,Hardin:1992mi,Hardin:1996bv}. These  are configurations of $N$ points on a sphere such that the average value of any polynomial of degree at most $t$ has the same average over the $N$ points as over the sphere. Thus, the $N$ points can be seen to give a representative average value of any polynomial of degree $t$ or lower.  It has been conjectured that a state is maximally unpolarized if and  only if its Majorana constellation is a spherical $t$-design~\cite{Crann:2010qd}. However, although the statement is true for some $t$-designs, such as those represented by the Platonic solids, the conjecture is not true in general~\cite{Bannai:2011pi}.

Nonetheless, there should be some connection between the number of points $N$ and the maximal degree $t$ for which a spherical $t$-design exists. The configurations that maximize $t$ for a given $N$ are called optimal designs, and in the following $t$ will denote the degree of an optimal $N$-point design. No analytical expression is known between $N$ and $t$: the number of points $N$ is at least proportional to $t^2$; whereas, for some orders $t$, the only known constructions have $N$ scaling proportionally to $t^3$.  As a function of $N$, the order $t$ is nonmonotonic. The current state of knowledge is summarized for
$1 \leq N \leq 100$ in~Ref.~\onlinecite{Hardin:1996bv}.

{In Ref.~\onlinecite{Bjork:2015ab} it has been conjectured that, for   a given $S$, the maximal order $M_{\mathrm{max}}$  for which  $\mathcal{A}_{M_\mathrm{max}}^{(S)}$ vanishes coincides with $t_{\mathrm{max}}$ in the corresponding spherical design. This is an intriguing connection, but further work is needed to support this.} 

{We also  note that an optimal $t$-design does not necessarily give a $t$th-order unpolarized state; this  underscores the mystery that the optimal $t$ and maximal $M$ always seem to be equal for any $N$. Another similarity between optimal spherical $t$-designs and the Kings is that the configurations are typically not unique, aside from those in the smallest dimensions.}

{The other related problem we  consider is the Thomson problem~\cite{Thomson:1904qp,Ashby:1986bk,Edmundson:1992uf,Melnyk:1977gm}, which consists of arranging $N$ identical point charges on the surface of a sphere so that the electrostatic potential energy of the configuration is minimized. The problem can be generalized to potential energies of the form $r^{-d}$, where $r$ is the Euclidean distance between the charges.  The case $d \rightarrow \infty$ is called the Tammes problem~\cite{Tammes:1930rc}. In contradistinction with the previous designs, the solution of the Thomson problem appears to be unique for every $S$~\cite{Erber:1991lq}.}

{To conclude, we mention that our $(2S+1)$-dimensional subspace that carries the irrep  of SU(2) can be considered  as the symmetric subspace of a system of $2S$ qubits. In this scenario, the Kings appear to be closely linked to other intriguing problems, such as maximally entangled symmetric states~\cite{Aulbach:2010jw,Giraud:2015oj} and $k$-maximally mixed states~\cite{Arnaud:2013hm,Goyeneche:2014so}.}

\subsection{Metrological power}

Following our earlier discussion of parameter estimation in Sec. \ref{sec:CV metrological power}, we can identify states that are the most and least sensitive for the detection of SU(2) operations. This sensitivity is characterized by the quantum Cram\'er-Rao bound \eqref{eq:qCRB} and depends only on the variance of a particular generator for any quantum state. 

The operation corresponding to a rotation by angle $\chi$ about axis $\mathbf{n}(\theta,\phi)$ is generated by {$\op{R}(\chi,\mathbf{n}) = \exp(i\chi \op{\mathbf{S}}\cdot \mathbf{n})$. Consequently, for pure states, an estimate of the rotation angle is appraised by the quantum Fisher information}
\begin{equation}
\Fish(\psi,\chi,\mathbf{n}) = 4 \, 
\Var_\psi (\op{\mathbf{S}} \cdot \mathbf{n}) \, .
\end{equation}

Considering a case in which information about the rotation axis $\mathbf{n}$ is irrelevant, we can average over all axes to yield
\begin{equation}
\overline{\Fish} (\psi) = \frac{1}{4\pi} 
\int_0^{2\pi}d\phi\,\int_0^{\pi}  d\theta \, \sin \theta
\Fish (\psi,\chi,\mathbf{n}) = 
\frac{4}{3} [\Var_\psi (\op{S}_x) + \Var_\psi (\op{S}_y) + \Var_\psi(\op{S}_z) ] = \frac{4}{3} \Delta^{2} \, .
\end{equation}
Per \eqref{eq:SU(2) total variance} this quantity is again minimized by the SU(2) coherent states; in contrast, it is  maximized by the first-order unpolarized states, {for which $\langle \op{\mathbf{S}} \rangle\sugg{_\psi} =\mathbf{0}$ and $\bar{\Fish}(\psi) = \tfrac{4}{3}S(S+1)$.} This is identical to the CV case, in which the greatest metrological power is found in states whose first order moments vanish and the least metrological power is found in the minimum-uncertainty states.

Just like with the {CV case}, it is insightful to average the variances of all estimates of the rotation angle $\chi$ by averaging the quantum Cram\'er-Rao bound:
\begin{equation}
{\overline{\Var}_\psi (\chi) \geq \frac{1}{4\pi}\int_0^{2\pi}d\phi\,\int_0^{\pi} d\theta \, \sin \theta \; 
\frac{1}{4 \, \Var_\psi (\op{\mathbf{S}}\cdot \mathbf{n})} \, \geq 
\frac{1}{\bar{\Fish}(\psi)} = \frac{3}{4\Delta^2} \, ,}
\end{equation}
{where we have again invoked Jensen's inequality. This is saturated if and only if} $\Fish(\psi,\chi,\mathbf{n})$ is independent of $\mathbf{n}$; again, this is only the case when the covariance matrix of the $\mathfrak{su}$(2) generators is isotropic $\Cov_\psi(\op{S}_i,\op{S}_j)\propto \delta_{i,j}$, which is equivalent to states being second-order unpolarized. We {thus conclude that the most powerful states for estimating rotation angles about arbitrary rotation axes are second-order unpolarized states~\cite{Martin:2020aa}.}

At the opposite extreme, coherent states lead to divergent lower bounds on the average variance that can be estimated for a parameter. Aligning the $z$-axis of the coordinate system $(\theta,\phi)$ with the coherent state $| \mathbf{n} \rangle = |{S,S} \rangle $, the covariances $\Cov_\psi (\op{S}_i,\op{S}_j)$ all vanish other than $\Cov_\psi(\op{S}_1,\op{S}_1)= \sugg{\Cov_\psi(\op{S}_2,\op{S}_2)}=\tfrac{S}{2}$. {For such a state, the averaged quantum Cram\'er-Rao bound diverges. \sugg{This may not be the only state that leads to divergent lower bounds (for instance, perhaps some of the squeezed ``intelligent states'' that generalize CV squeezed states \cite{Mahler:2010fk} perform, on average, equally to coherent states); as with the CV case, information from the first moment alone may hide deficiencies in metrological power.} In reality, of course, one can assume that the angle $\chi\in\left[0,2\pi\right)$, which even for a uniform distribution \sugg{has its variance bounded} from above by $\pi^2/3$. The conclusion, in both groups studied here, is that coherent states have the least metrological power and second-order unpolarized states have the greatest metrological power.} 

In light of the previous discussions {we predict that states whose higher-order moments are also isotropic, i.e., higher-order unpolarized states, are even more useful for some metrological tasks.}

\section{Concluding remarks}

Extremal principles provide a global and elegant formulation of  physical phenomena. They probably represent the best example of economy of thought in physics~\cite{Born:1939aa}. We have discussed several of these principles and how they can be used to determine the most and the least quantum states, without trying to quantify quantumness.

Although these principles are universal, their particular application depends on the system under investigation. We have discussed two significantly different scenarios: continuous-variable and discrete-variable information.
In the first case, the conclusions for all our principles seem to be very similar: coherent states are the least quantum states and number states are the most quantum ones. However, the harmonic oscillator, which lies at the heart of the CV approach, is too special; very often, different notions yield the same results for this relevant model. The very concept of what makes a state coherent is a good example of this~\cite{Perelomov:1986ly}.

For the case of DV systems, things are different: whereas the least quantum states are still coherent states, the most quantum ones depend on the extremal principle at hand. In some particular dimensions, they coincide, but they differ for others. {The differences  become  especially dramatic as $S \rightarrow \infty$, which goes against the naive expectation that in that limit one should recover the CV world. Much more work must be done to understand properly this transition. } 

We have discussed how these quantum states  may be exploited to beat classical bounds. This elevates the notion of quantumness from something that is of purely fundamental interest to a resource with practical utility.

We believe that a proper understanding of the key concepts concerning quantumness is a necessary step for harnessing all the magical weirdness of the quantum world. 

\section*{Ackowledgments}
{We are indebted to I. Bengtsson, G. Bj\"{o}rk, \sugg{A. Hertz,} P. de la Hoz,  H. Jeong, \sugg{W. Vogel}, and K.  \.{Z}yczkowski for discussions.   We acknowledge financial support from the Mexican CONACYT (Grant 254127), the Spanish MINECO (Grant PGC2018-099183-B-I00), \sugg{the European Union Horizon 2020 (Grant ApresSF),  the Foundation for Polish Science (IRAP project, ICTQT, contract no. 2018/MAB/5, co-financed by EU within Smart Growth Operational Programme), and a Mega-grant of the Ministry of Education and Science of the Russian Federation (Contract No. 14.W03.31.0032).} AZG acknowledges funding from an NSERC Discovery Award Fund, an NSERC Alexander Graham Bell Scholarship, the Walter C. Sumner Foundation, the Lachlan Gilchrist Fellowship Fund, a Michael Smith Foreign Study Supplement, and Mitacs Globalink.

\section*{Data availability}

Data sharing is not applicable to this article as no new data were created or analyzed in this study.

%

\end{document}